\documentclass[a4paper,fleqn,twocolumn,margin=1in]{aastex61} 
\pdfoutput=1
\usepackage{float}
\usepackage{graphicx} 
\usepackage[font={small,it}]{caption} 
\usepackage{ae,aecompl}

\usepackage{enumitem} 
\usepackage{amssymb} 
\usepackage{amsmath} 
\usepackage[lofdepth,lotdepth,caption=false]{subfig}

\def\apjl{\rm ApJL} 
\def\apjs{\rm ApJS}

\def\aap{\rm AAP}

\def\gax{\mathrel{\raise.3ex\hbox{$>$}\mkern-14mu\lower0.6ex\hbox{$\sim$}}} 
\def\lax{\mathrel{\raise.3ex\hbox{$<$}\mkern-14mu\lower0.6ex\hbox{$\sim$}}} 
\def\gtorder{\mathrel{\raise.3ex\hbox{$>$}\mkern-14mu 
             \lower0.6ex\hbox{$\sim$}}} 
\def\ltorder{\mathrel{\raise.3ex\hbox{$<$}\mkern-14mu 
             \lower0.6ex\hbox{$\sim$}}} 
\def\plxtgas{\varpi_{\mathrm{TGAS}}}
\def\plxastero{\varpi_{\mathrm{astero}}}

\def\plxtgasi{\varpi_{\mathrm{TGAS, i}}}
\def\plxasteroi{\varpi_{\mathrm{astero, i}}}
\def\plxtgasj{\varpi_{\mathrm{TGAS, j}}}
\def\plxasteroj{\varpi_{\mathrm{astero, j}}}
\def\muhz{\mu\mathrm{Hz}}
\def\numax{\nu_{\mathrm{max}}}
\def\dnu{\Delta \nu}

\def\teff{T_{\mathrm{eff}}}
\def\teffapogee{T_{\mathrm{eff, APOGEE}}}
\def\teffirfm{T_{\mathrm{eff, IRFM}}}
\def\feh{\mathrm{[Fe/H]}}
\def\teffsun{T_{\mathrm{eff,} \odot}}

\def\fbol{F_{\mathrm{bol}}}

\def\rsun{R_{\odot}}
\def\deg{^{\circ}}
\def\msun{M_{\odot}}
\def\numaxsun{\nu_{\mathrm{max,} \odot}}
\def\dnusun{\Delta\nu_{\odot}}
\def\logg{\log~g}
\def\loggsun{\log~g_{\odot}}

\begin{document}
\title{Evidence for spatially-correlated {\it Gaia} parallax errors in
the {\it Kepler} field}
%\shorttitle{Evidence for spatially-correlated {\it Gaia} parallax errors}
%\shortauthors{Zinn et al.}
\author{Joel C. Zinn}
\affiliation{Department of Astronomy, The Ohio State University, 140 West
  18th Avenue, Columbus OH 43210, USA}
\author{Daniel Huber}
\affiliation{ Institute for Astronomy, University of Hawai`i, 2680 Woodlawn Drive, Honolulu, HI 96822, USA}
\affiliation{Sydney Institute for Astronomy (SIfA), School of Physics,
University of Sydney, NSW 2006, Australia}
\affiliation{Stellar Astrophysics Centre, Department of Physics and
Astronomy, Aarhus University, Ny Munkegade 120, DK-8000
Aarhus C, Denmark}
\affiliation{SETI Institute, 189 Bernardo Avenue, Mountain View, CA
  94043, USA}
\author{Marc H. Pinsonneault}
\affiliation{Department of Astronomy, The Ohio State University, 140 West
  18th Avenue, Columbus OH 43210, USA}
\author{Dennis Stello}
\affiliation{School of Physics, University of New South Whales, Barker
  Street, Sydney, NSW 2052, Australia}
\affiliation{Sydney Institute for Astronomy (SIfA), School of Physics,
  University of Sydney, NSW 2006, Australia}
\affiliation{Stellar Astrophysics Centre, Department of Physics and
Astronomy, Aarhus University, Ny Munkegade 120, DK-8000
Aarhus C, Denmark}

\correspondingauthor{Joel C. Zinn}
\email{zinn.44@osu.edu}

\accepted{25 June 2017}
\submitjournal{The Astrophysical Journal}

\begin{abstract}
We present evidence for a spatially-dependent
systematic error in the first data release of {\it Gaia} parallaxes based on comparisons to
asteroseismic parallaxes in the {\it Kepler} field, and present a parametrized model of the
angular dependence of these systematics. We report an error of $0.059^{+0.004}_{-0.004}$mas on scales
of $0.3$deg, which decreases for larger scales to become $0.011^{+0.006}_{-0.004}$mas at
$8$deg. % STAT
This is consistent with the $\sim 2 \%$ zeropoint offset for the whole
sample discussed by Huber et al., and is compatible with the effect
predicted by the {\it Gaia} team. Our results are robust to dust
prescriptions and choices in temperature scales used to calculate
asteroseismic parallaxes. We also do not find evidence for significant
differences in the signal
when using red clump versus red giant stars. Our
approach allows us to quantify and map the correlations in an
astrophysically interesting field, resulting in a parametrized model of
the spatial systematics that can be used to construct a covariance matrix
for any work that relies upon TGAS parallaxes.
\end{abstract}

\section{Introduction}

The {\it Gaia} mission is expected to provide positions,
parallaxes, and proper motions for a billion objects, with precisions
of $\sim 20$ micro-arcseconds ($\mu$as) for stars down to $15^{\mathrm{th}}$
magnitude \citep{Gaia2016b}. Though the final data release is scheduled for
2022\footnote{\href{http://www.cosmos.esa.int/web/Gaia/release}{http://www.cosmos.esa.int/web/Gaia/release}}, positions,
parallaxes, and proper motions for $2$ million
stars common to Tycho-2 \citep{hog+2000} and {\it Gaia} have been released as part
of Data Release 1 (DR1) \citep{Gaia2016a}. By using positions from Tycho-2
as { priors on} the astrometric solution, \cite{michalik_lindegren&hobbs2015} demonstrated
that sub-milliarcsecond accuracy could be achieved, resulting in the
Tycho-{\it Gaia} Astrometric Solution (TGAS). Though the
statistical errors may even be smaller than the $0.3$mas reported in DR1
\citep[see][]{gould_kollmeier_sesar2016},
systematic errors are expected to exist at the level of up to
$0.3$mas \citep{lindegren+2016}. Various instrumental and modeling
effects that may account for the systematic errors are explored in
\cite{lindegren+2016}, including the known chromaticity of the CCDs,
inadequate temporal resolution of the satellite attitude model,
and so-called `micro-clanks' due to mechanical jitter.

In this work, we compute asteroseismic parallaxes for
more than 1000 red giants in the $10\deg \times10\deg$ {\it Kepler} field
of view for comparison to TGAS parallaxes. Thanks to the
order-of-magnitude better precision of asteroseismic parallaxes for
red giants, and
the high stellar density of the {\it Kepler} field, we are able to investigate the presence of
correlated errors in TGAS parallaxes on degree and sub-degree scales in an effort
to quantify the expected systematic spatial errors in TGAS parallax.

To date, other comparisons of the asteroseismic parallax scale to the TGAS
parallax scale
have considered global offsets --- i.e., non--spatially-dependent
differences. \cite{deridder+2016},
for instance, found good agreement between the two scales for a sample
of
22 dwarfs and sub-dwarfs, but significant differences among 938 red
giants. Huber
et al. (in press) suggest that a global offset is partially mitigated when using
a hotter temperature scale such as the infrared
flux method temperature scale, and that radii inferred from TGAS
parallaxes are
consistent with asteroseismic radii to within $5\%$ between $0.8$--$8 \rsun$.

Compared to other parallax scales, the TGAS parallax scale exhibits a
fractional offset. \cite{davies+2017}, for example,
attributed
the red giant asteroseismic parallax offset from \cite{deridder+2016}
to errors in
the TGAS parallax scale by comparing TGAS parallaxes to red clump distances. Their
suggested correction agrees
for $\varpi \gtrsim 1.6$ with that of \cite{stassun&torres2016a}, who
compared TGAS parallaxes to
parallaxes from eclipsing binaries. Huber et al. (in press) found that
these offsets are too large, and can be partially attributed to a
too cool temperature scale, based on a larger sample of stars spanning
from the main sequence to the red giant branch. At larger distances,
\cite{sesar+2017} found no evidence for a global offset when comparing to RR Lyrae parallaxes at a median parallax of $0.8$mas, and
neither did \cite{casertano+2017}
when looking at Cepheid parallaxes. And at the smallest distances,
\cite{jao+2016} found
evidence for a correction consistent with that of
\cite{stassun&torres2016a} (amounting to $\approx 0.2$mas
in the sense that TGAS overestimates distances) when compared to
trigonometric parallaxes
for 612 dwarfs at distances of less than $100$pc.

To date, two studies independent of the {\it Gaia} team have mentioned
possible spatial dependencies in TGAS parallax
scales. \cite{casertano+2017} found mild evidence for
spatially-correlated TGAS parallaxes at the level of $19 \pm34\mu$as
on scales
less than $10$deg using Cepheids, while \cite{jao+2016} reported a
North-South
ecliptic hemisphere difference in trigonometric and TGAS
parallaxes. Despite
the thorough investigation of the quantitative and qualitative
existence of such systematic errors in \cite{lindegren+2016}, the
precise characterization of
spatially-dependent systematics in terms of a functional form and/or a
characteristic scale at which the $0.3$mas systematic error applies
was not released for DR1. Our exercise, then, is to identify and to
characterize any spatial correlation of
parallax errors in {\it Gaia} DR1.

The structure of this paper is as follows: we describe the provenance
of and basic calibrations of the observables used to compute
asteroseismic parallax in \S\ref{sec:data}. In \S\ref{sec:methods}, we
detail how we compute asteroseismic parallaxes, treatment of
statistical errors therein, and how we test for the presence of
spatially-correlated offsets between asteroseismic TGAS parallaxes. We summarize our main
findings in \S\ref{sec:results}, discuss potential caveats to those
findings in \S\ref{sec:discussion}, and conclude in \S\ref{sec:conclusion}.

\section{Data}
\label{sec:data}
Quantifying any systematic errors in {\it Gaia} parallaxes requires an
independent and unbiased set of parallaxes to compare to the TGAS
values. The {\it Gaia} team validated their parallaxes against {\it
  Hipparcos} parallaxes, which revealed the presence of systematic
offsets \citep{lindegren+2016}. We attempt to present a complementary
treatment of potential errors in the TGAS parallax scale for two main
reasons. First, the {\it Hipparcos} parallaxes themselves have
spatially-correlated errors (see references in
\S\ref{sec:discussion}), which limits their usefulness when used as a
validation set. More critically, a detailed model of the spatial
correlations of TGAS parallax errors has not yet been published, which would be crucial to proper treatment of errors in work using TGAS between now and April 2018, when the next {\it Gaia} data release is scheduled.

Our validation set consists of parallaxes of red giants in the {\it
  Kepler} field of view that have spectroscopic metallicities and
asteroseismic data, which permits us to infer effective temperature,
radii, and, by extension, luminosities. Adding reddening and
bolometric flux information then yields { distances and hence
parallaxes}. The resulting asteroseismic parallaxes have statistical
errors an order-of-magnitude smaller than those in TGAS, and hence permit
a strong test of spatially-correlated offsets in TGAS versus
asteroseismic parallax scales.

Our sample consists of over 1000 red giants spread across the
$\sim 100$ sq. deg. {\it Kepler} field of view, which means that we
can probe systematic parallax offsets on scales less
than a degree. This is the scale where \cite{lindegren+2016} indicate systematic errors in the TGAS parallaxes are expected to be the largest.

The basis for our sample are TGAS stars that are also listed as
asteroseismic giants in the APOGEE-Kepler Asteroseismic Science
Consortium catalogue \citep[APOKASC;][]{pinsonneault+2014}, which combines infrared spectroscopic data from Data Release 13 of the The Apache Point Observatory Galactic Evolution
Experiment (APOGEE) \citep{zasowski+2013, majewski+2015} with
asteroseismic data from the {\it Kepler} mission
\citep{borucki+2010}. We now discuss the provenance of spectroscopic metallicities, asteroseismic parameters, reddenings, and photometry in turn.

APOGEE temperatures, $\teffapogee$, and metallicities, $\feh$, are taken from the Thirteenth
Data Release of the Sloan Digital Sky Survey
\citep[DR13; SDSS;][]{SDSS+2016}, and are corrected according to the
metallicity-dependent term recommended in the DR13
documentation\footnote{\href{http://www.sdss.org/dr13/irspec/parameters/}{http://www.sdss.org/dr13/irspec/parameters/}}. Global
asteroseismic parameters $\numax$ and $\dnu$ --- which may be mapped
into stellar { radii} --- were adopted from
the SYD pipeline \citep{huber+2009} values in version 3.6.5 of the
APOKASC catalogue (Pinsonneault et al., in prep.).

Because there is evidence that asteroseismic radii have evolutionary
state--dependent systematics \citep[e.g.,][]{miglio+2012}, we divide
the TGAS-APOKASC giant sample into red giant branch (RGB) and red
clump (RC) sub-samples, to assess any differences in TGAS-asteroseismic parallax offsets as a function of evolutionary state. Evolutionary state information is compiled from
the asteroseismic
classifications of \cite{stello+2013} or \cite{mosser+2014} (Elsworth
et al., in prep.).

\begin{figure}[htb!]
\includegraphics[width=0.5\textwidth]{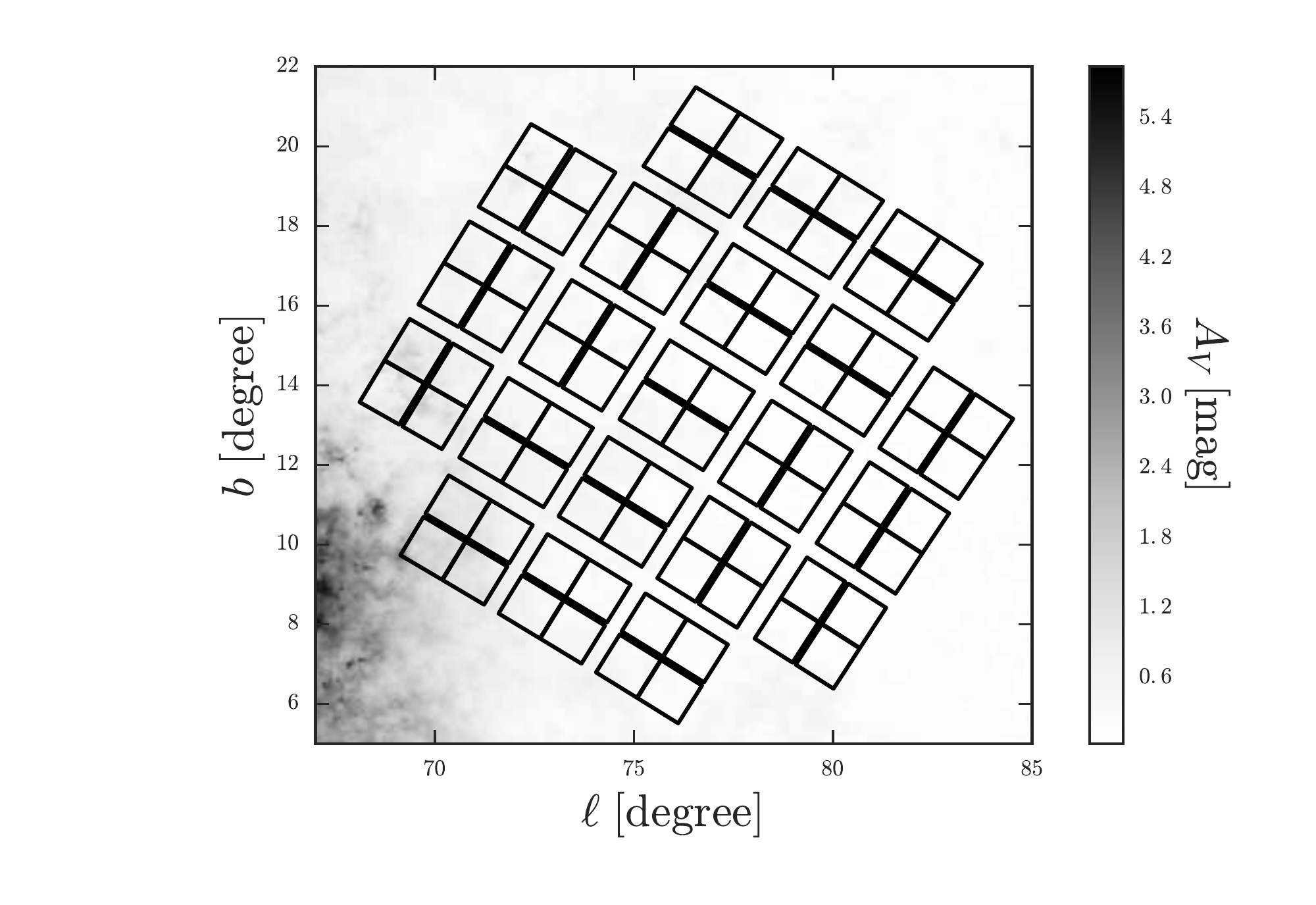}
\caption{We use a { three-dimensional} dust map from \protect\cite{green+2015}, as implemented in \texttt{mwdust} \protect\citep{bovy+2016}. Shown here is $A_V$ in the region of
  the {\it Kepler} field of
  view, in Galactic coordinates. Choosing to include or not the higher extinction region $\ell \lesssim 73^{\circ}$ does not eliminate spatially-correlated offsets between asteroseismic and TGAS parallaxes.}
\label{fig:dust}
\end{figure}

Extinction corrections (described in \S\ref{sec:plxastero}) are made
using the three-dimensional dust map of \protect\cite{green+2015}, as
implemented in \texttt{mwdust}\footnote{\href{https://github.com/jobovy/mwdust}{https://github.com/jobovy/mwdust}} \citep{bovy+2016}. The $A_V$ extinction for the
{\it Kepler} field of view is shown in Figure~\ref{fig:dust}.

We opt to calculate an effective temperature and bolometric flux using
the InfraRed Flux Method (IRFM), as implemented in \cite{GHB09}, which
was used to set the APOGEE effective temperature scale. For this purpose, we use near-infrared photometry in the $J$, $H$, and $K_{\mathrm{s}}$ bands from the Two Micron All Sky Survey
\citep[2MASS;][]{skrutskie+2006}. Visual photometry is also required, which we derive from SDSS $g$ and $r$ photometry. We choose to convert these magnitudes to Johnson $B$
and $V$ according to Lupton
(2005)\footnote{\href{https://www.sdss3.org/dr10/algorithms/sdssUBVRITransform.php}{https://www.sdss3.org/dr10/algorithms/sdssUBVRITransform.php}}
rather than use Tycho $B$ and $V$. In doing so, the resulting visual
photometry has less scatter than Tycho $B$ and $V$. Furthermore, the
$griz$ photometry from the {\it Kepler} Input Catalogue
\citep[KIC;][]{brown+2011}, as re-calibrated to be on the SDSS scale
by \cite{pinsonneault+2012a}, is consistent with the 2MASS infrared
photometry temperature scale for cool stars
\citep[see][]{pinsonneault+2012a}.

{ The requirement
that our sample of stars have $grJHK_{\mathrm{s}}$,
$\numax$, $\dnu$, $\teffapogee$, $\feh$, RGB or RC evolutionary state
classifications, and {\it Gaia} DR1 parallaxes ($\plxtgas$) yields a base sample of 1592 giants}. %STAT

\subsection{Quality cuts}
We omit stars known to be members of
NGC6791 and NGC6819, as giants residing in these clusters could bias
measurements of spatially-correlated quantities. 

Comparisons by \cite{gould_kollmeier_sesar2016} of TGAS parallaxes to
RR Lyrae parallaxes indicated that DR1 TGAS parallax statistical errors are
inflated by $\sim 30\%$. Because the calculation of a spatially-correlated TGAS-asteroseismic
parallax offset will be more robust with a proper treatment of the
statistical errors, we modify those for the TGAS parallaxes according to their
prescription. Reducing statistical errors in this way does not introduce a spatially-correlated, systematic offset of the sort we present in this work.

{ Finally, TGAS parallaxes are required to have a signal-to-noise
  ratio greater than $1.6$ (see \S\ref{sec:sample}). %STAT

The above quality cuts yield a total of 1392 giants, which comprise %STAT
the final TGAS-APOKASC sample used in
the following.}

\section{Methods}
\label{sec:methods}
\subsection{Asteroseismic parallax}
\label{sec:plxastero}
Estimating errors in TGAS parallaxes requires an independent distance measure. Apart from the moving group or
parallax methods, distance estimates of stars will require an estimate
of stellar luminosity and its bolometric flux. For our purposes, we
use asteroseismology to determine stellar luminosity and the infrared
flux method to determine a bolometric flux, which are combined to
yield a parallax/distance. As the following overview will show,
asteroseismology effectively provides a radius, which in combination
with an effective temperature of the star, will determine its
luminosity via the Stefan-Boltzmann equation; combined with the
bolometric flux of the star, one can determine its distance.

In this work, we estimate stellar radius by way of two complementary
scaling relations involving two different asteroseismic observables: $\numax$ (roughly the frequency at which the largest-amplitude
acoustic modes occur) and $\dnu$, the separation between acoustic
modes of the same spherical harmonic degree, $\ell$, but differing by one
radial order number, $n$.

It is well-established \citep[see,
  e.g.,][]{tassoul1980,christensendalsgaard1993} that $\dnu$ is
related to the mean density of a star via a scaling relation,
assuming homologous behavior between the Sun and a given star, of the form
\begin{equation}
\frac{\dnu}{\dnusun} \approx \sqrt{\frac{M/\msun}{(R/\rsun)^3}}.
\label{eq:scaling1}
\end{equation}

Similarly, the frequency of maximum acoustic power, $\numax$, has been
found to scale as the acoustic cutoff frequency \citep{brown+1991,kjeldsen&bedding1995,chaplin+2008}, i.e., as 

\begin{equation}
\frac{\numax}{\numaxsun} \approx \frac{M/\msun}{(R/\rsun)^2\sqrt{(\teff/\teffsun)}}.
\label{eq:scaling2}
\end{equation}

We can combine these two relations to yield an estimate of the radius, $R$,
of the star:
\begin{equation}
(R/\rsun) \approx (\numax/\numaxsun)(\dnu/\dnusun)^{-2} (\teff/\teffsun)^{1/2}.
\label{eq:scaling3}
\end{equation}

With a temperature and the radius, we can compute a luminosity and
thus a luminosity distance/parallax, provided we know the bolometric
flux, $\fbol$:

\begin{align}
\plxastero = \frac{\sqrt{\fbol}}{R \sqrt{\sigma_{\mathrm{SB}}}
  \teff^2},
\label{eq:plxastero}
\end{align}
with $\sigma_{\mathrm{SB}}$ being the Stefan-Boltzmann constant. We
adopt solar values consistent with those of \cite{huber+2009}: $\numaxsun=3090 \muhz$; $\dnusun =
135.1 \muhz$; $\teffsun = 5777$K; and $\loggsun = 4.438$ \citep{mamajek+2015}.

\begin{figure}[htb!]
\includegraphics[width=0.5\textwidth]{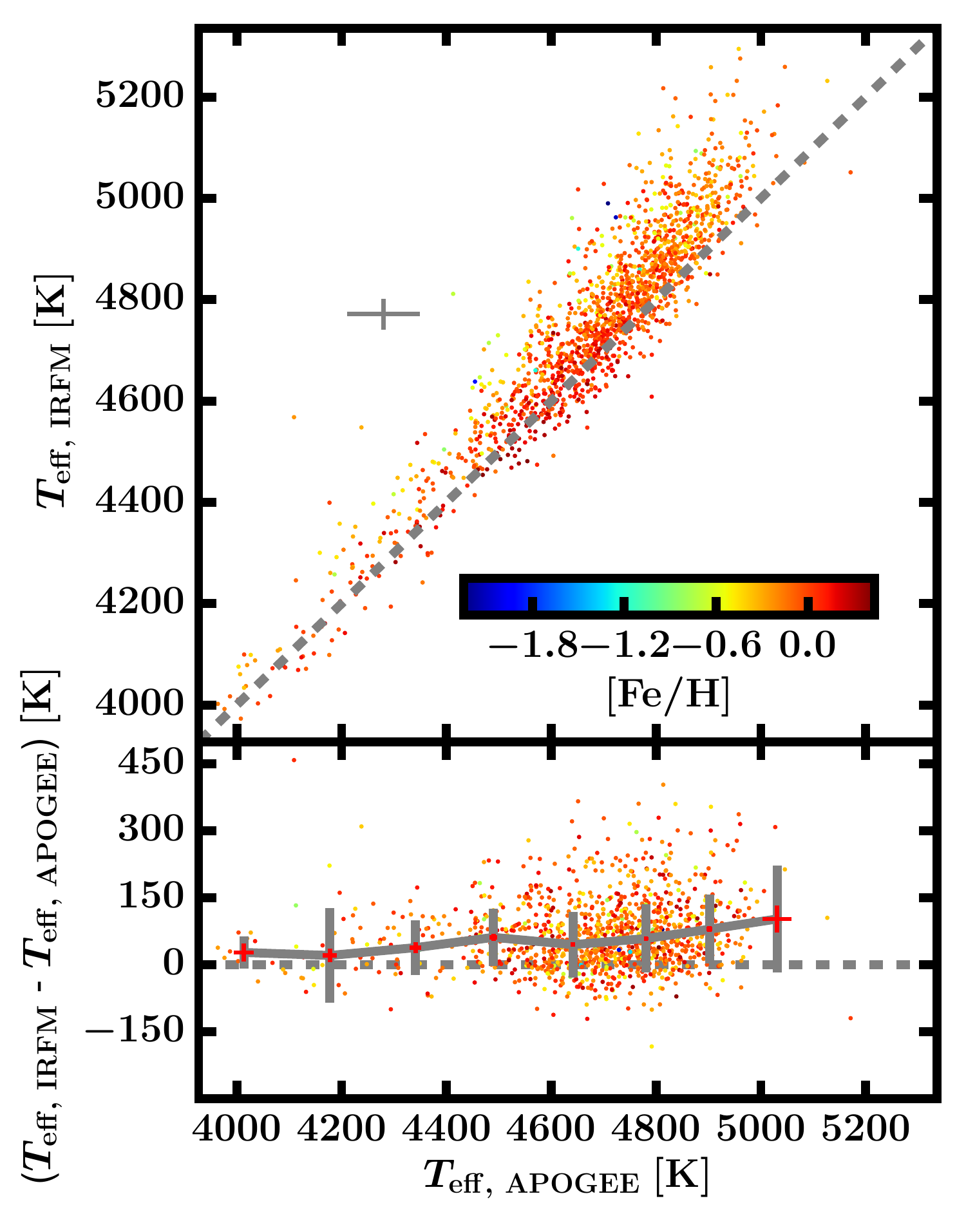}
\caption{APOGEE and IRFM temperature scales show a systematic offset
  that is temperature-dependent. We find that $\teffirfm$ results in
  parallaxes more consistent with those from TGAS, and use this
  temperature scale instead of $\teffapogee$ throughout the
  paper. { Grey dashed lines show one-to-one relations.} Median errors on both quantities are shown by the error bar
  in the top panel. The
bottom panel shows a { binned} median of the temperature
difference (grey curve), with grey error bars representing the standard
deviation of the difference within each bin and red error bars
representing the statistical error { on the median} within each bin. The IRFM temperature is hotter than APOGEE spectroscopic
temperatures by $65 \pm 13$K, on average. }% STAT
\label{fig:apogee_v_irfm}
\end{figure}

With a radius from asteroseismology, we turn to the bolometric flux and effective temperature,
which we infer from the infrared flux method (IRFM) using $BVJHK_{\mathrm{s}}$
photometry, according to \cite{GHB09}. Calculating an effective
temperature using the IRFM allows us to self-consistently estimate the reddening (and
hence extinction) to each star, which is necessary to achieve a
correct distance measure. The basic approach is to simultaneously fit a star's spectral energy distribution
from the optical to the infrared, taking advantage of the
insensitivity of infrared stellar emission on effective
temperature. First, the observed infrared flux is compared to the infrared
flux for a model atmosphere, yielding an angular diameter. Next, a bolometric flux is computed by combining other photometric information (e.g., optical) with infrared photometry, based on an assumed stellar atmosphere model. Finally, a temperature is determined by using the previously computed bolometric flux and angular diameter. The bolometric flux and temperature results converge iteratively. Since the IRFM requires stellar
atmosphere lookups as a function of $\feh$, $\logg$, and $\teff$, we
implement the IRFM using guesses for these quantities from APOGEE. For the whole process,
we assume a fixed metallicity { from APOGEE, $\feh$}. Our initial guess
for $\teff$ is $\teffapogee$; our initial guess for $\logg$ is calculated from
Equation~\ref{eq:scaling2} using $\teffapogee$. An IRFM temperature and
bolometric flux are then computed iteratively, as described in
\cite{GHB09}. The resulting IRFM temperature, $\teffirfm$, is used to compute a new
$\logg$, and the bolometric flux is used via
Equation~\ref{eq:plxastero} to compute an asteroseismic
distance/parallax, $\plxastero$. An extinction for each band is then computed using the
three-dimensional dust map of \cite{green+2015} using \texttt{mwdust} \citep{bovy+2016}, with which we correct the $JHK_{\mathrm{s}}$
photometry, yielding dust--de-extincted $J_0$, $H_0$, and
$K_{\mathrm{s,\ 0}}$. $B_0$ and $V_0$ are computed by transforming corrected
$g$ and $r$ magnitudes, $g_0$ and $r_0$. This corrected photometry is
then used in subsequent iterations to compute the bolometric flux and
temperature, and the process is repeated until convergence in the
asteroseismic parallax. We compute uncertainties on the derived quantities $B_0$,
$V_0$, { $J_0$, $H_0$, $K_{\mathrm{s,\ 0}}$,} $A_V$, $\logg$, $\teffirfm$, and $\plxastero$ by repeating the
iterative process, perturbing the observable quantities
$g$, $r$, $J$, $H$, $K_{\mathrm{s}}$, $\numax$, $\dnu$, $\teffapogee$, and $\feh$
based on their statistical errors{ , and imposing a minimum uncertainty
of $0.08$mag for $A_V$ to account for variations in $R_V$ within the
{\it Kepler} field and for line-of-sight variations below the
resolution of the \cite{green+2015} dust map ($\sim 0.05$deg)}.  Resulting IRFM temperatures are
shown in Figure~\ref{fig:apogee_v_irfm}. When
compared to APOGEE spectroscopic temperatures, the IRFM temperatures
are on average $\sim 70$K
hotter. As we have found in Huber et al. (in press), the IRFM
temperature scale results in a smaller global offset between asteroseismic and
TGAS parallaxes than when using, e.g., spectroscopic temperatures from
APOGEE.

\begin{figure}[htb!]
\centering
\includegraphics[width=0.5\textwidth]{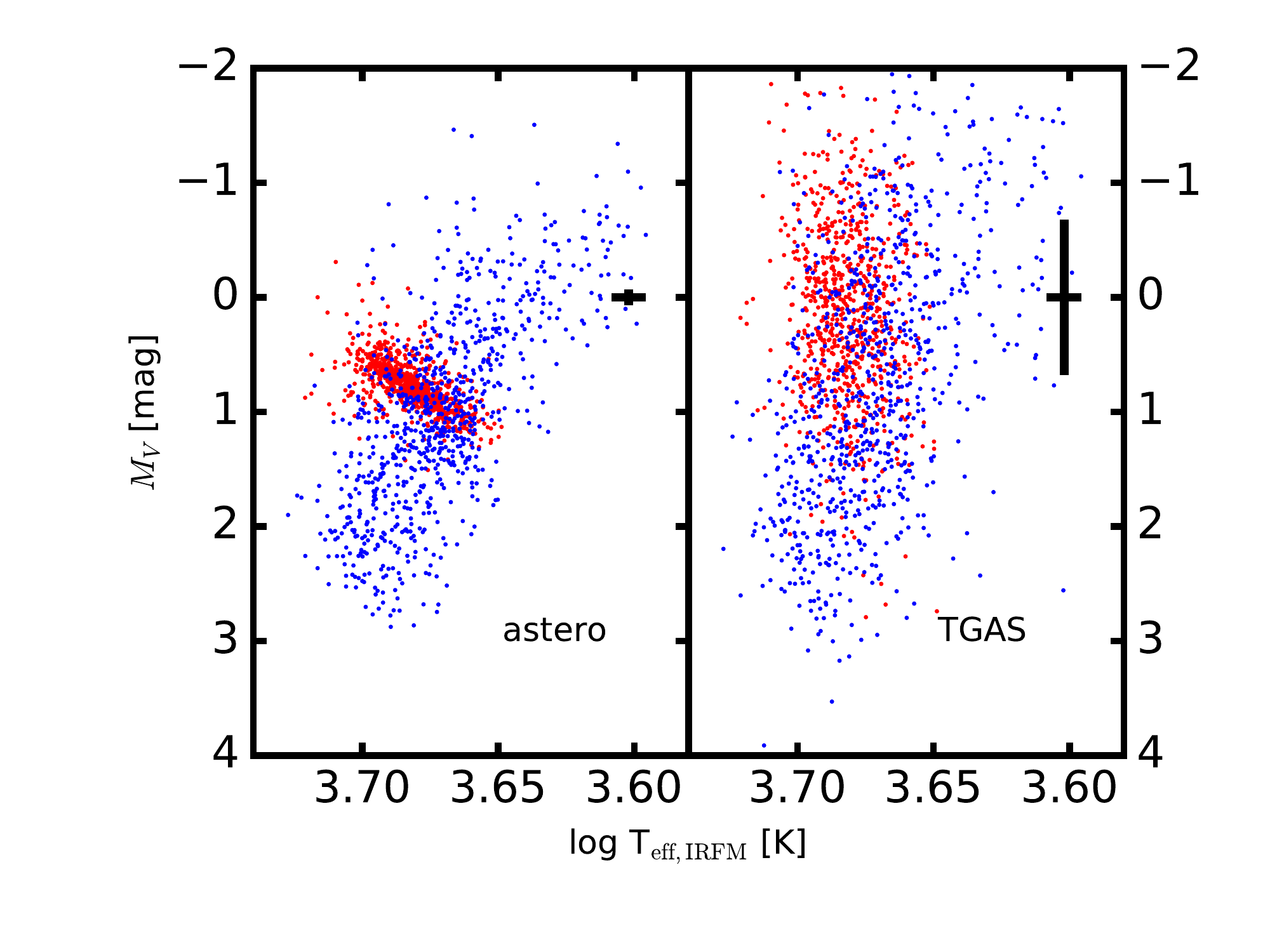}
\caption{The absolute
  magnitude in these Hertzsprung-Russell diagrams for the
  TGAS-ASPOKASC sample are computed via the IRFM (see text), and either an
  asteroseismic parallax (left) or a TGAS parallax (right). Median
  error bars are shown in black. Red clump stars are shown in red and
  red giant branch stars in blue.} 
\label{fig:hr}
\end{figure}

\begin{figure}[htb!]
\centering
\includegraphics[width=0.5\textwidth]{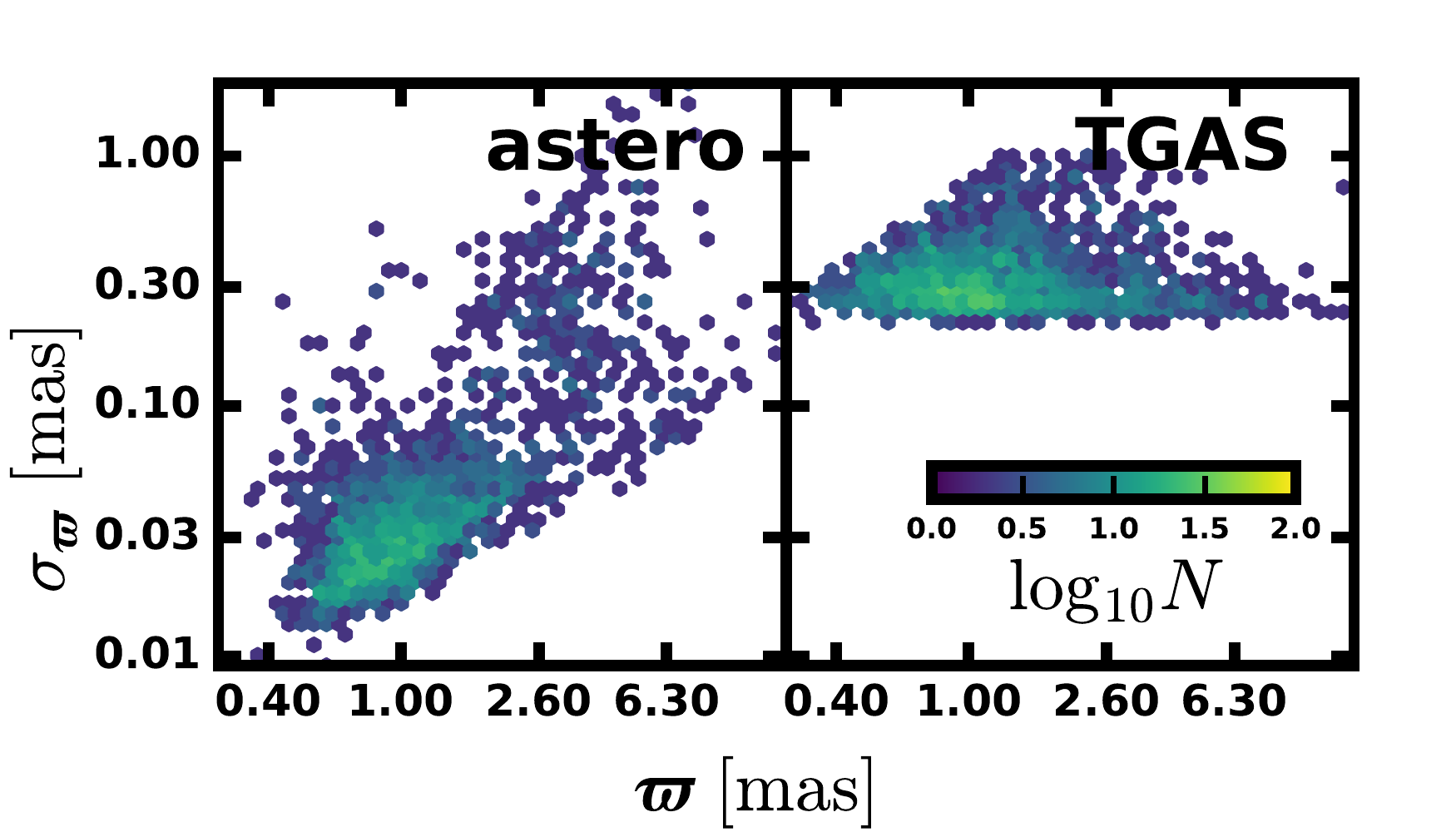}
\caption{{ A two-dimensional histogram of statistical errors on parallax
versus parallax for asteroseismic parallaxes derived in this work
(left) and for TGAS parallaxes (right). The former are an
order-of-magnitude more precise than the latter, which makes the
TGAS-APOKASC sample in this work a powerful 
calibrator for investigating systematic errors in
TGAS parallaxes.}}
\label{fig:errs}
\end{figure}

{ For purposes of illustration, Hertzsprung-Russell diagrams are
  constructed in Figure~\ref{fig:hr} from asteroseismic and TGAS
  parallaxes in combination with de-extincted V-band magnitudes, $V_0$. We employ an
  exponentially decreasing space density prior with a scale length of
  $1.35$kpc \citep{bailer_jones2015,astraatmadja&bailer_jones2016} for
  the conversion of parallax to distance.} The red clump is
particularly sharp using asteroseismic parallaxes compared to the
spread of red clump luminosities assuming TGAS parallaxes. {  Figure~\ref{fig:errs} demonstrates
  the relative precision of asteroseismic and TGAS parallaxes as a
  two-dimensional histogram of statistical parallax error versus
  parallax for the TGAS-APOKASC giant sample; the median uncertainty
  for asteroseismic parallaxes is $0.03$mas and the median uncertainty
  for TGAS parallaxes is $0.3$mas.} Clearly,  % STAT
the error budget in the comparisons between the scales is dominated
by TGAS parallax uncertainties. 

\begin{figure}[htb!]

\centering
\includegraphics[width=0.5\textwidth]{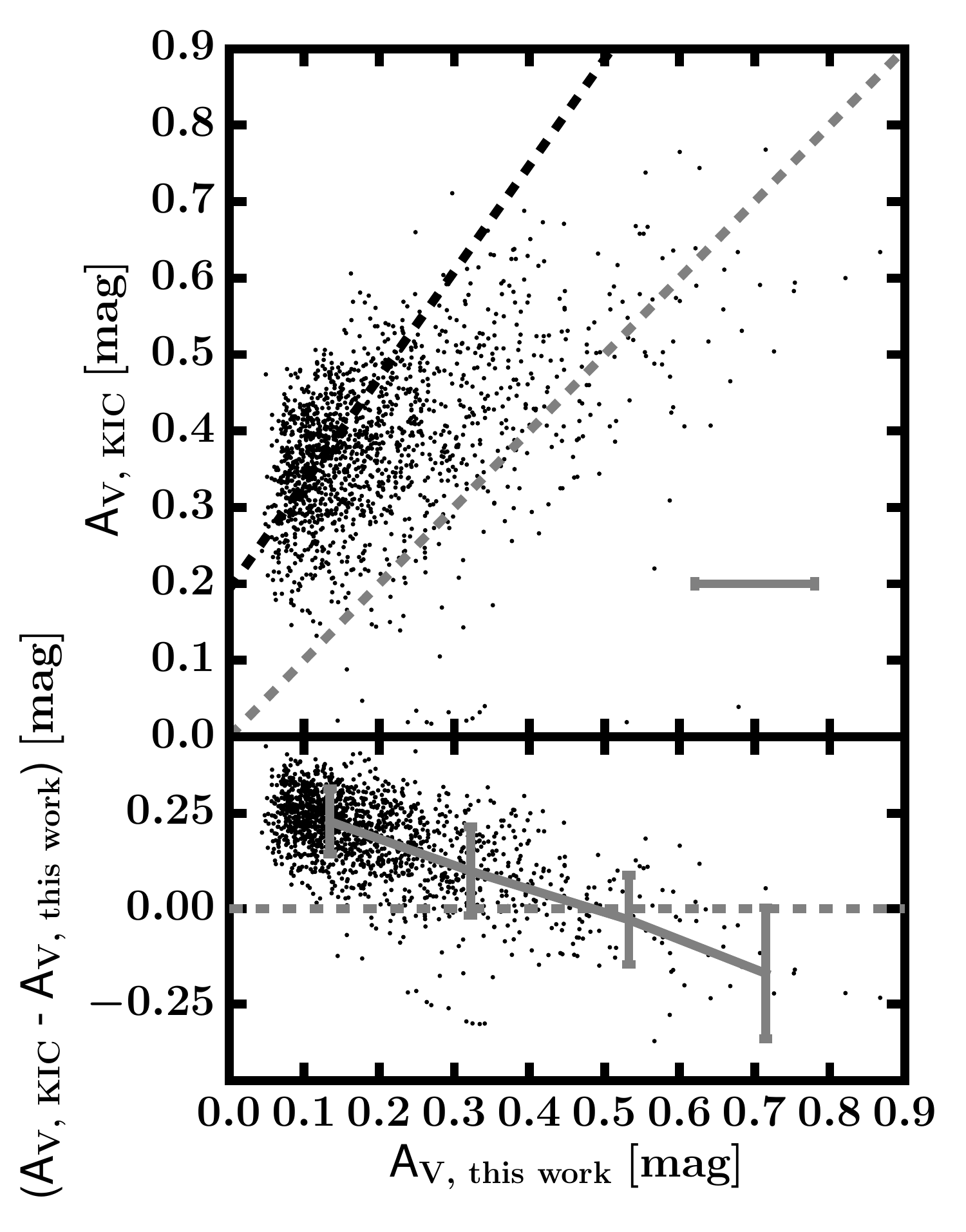}

\caption{{ Comparison of our $A_V$ to the KIC estimates
  \protect\citep{brown+2011}, $A_\mathrm{V,\ KIC}$. Grey dashed lines
  show one-to-one relations. The grey error
  bar in the top panel indicates the median uncertainty on $A_V$ for
  our derived extinctions. Note that
  uncertainties on extinctions are not reported for the KIC, and that
  a minimum uncertainty on our derived extinctions of 0.08mag is
  imposed. The black line is the
  relation between visual extinctions derived in \cite{rodrigues+2014}
  and KIC extinctions. The bottom panel
  shows a binned median, with grey error bars representing the scatter in each bin.}}
\label{fig:ext}
\end{figure}
\subsection{Extinction}
Extinction values are computed as a result of the iterative procedure
described in \S\ref{sec:plxastero}, which we have compared to
extinctions from the Kepler Input Catalogue
\citep[KIC;][]{brown+2011}. Previous studies suggest that the KIC
extinctions are over-estimated \citep{rodrigues+2014,zasowski+2015},
and we also find that our extinction values are smaller than those in
the KIC. { Figure~\ref{fig:ext} shows that the offset between our
derived extinctions and KIC extinctions is comparable to to the offset when checking against
Bayesian extinction estimates of APOKASC giants in
\cite{rodrigues+2014}, who found $A_\mathrm{V} = (0.721 \pm 0.015)
A_{\mathrm{V,\ KIC}} - (0.139 \pm 0.007)$. This relation is plotted as
a black dashed line on
top of our extinctions.} Our extinctions based on
\cite{green+2015} dust maps also
compare well to the extinctions derived from grid-based modeling in
Huber et al. (in press). \texttt{mwdust} offers several dust maps, and our result does not significantly change if using the \cite{green+2015} map or a combination of individual maps from  \protect\cite{marshall+2006}, \protect\cite{green+2015}, and
  \protect\cite{drimmel_cabrera_labers_lopez_corredoira2003}, as synthesized
  by \protect\cite{bovy+2016}.

\begin{figure*}[htb!]
\subfloat{
\centering
\includegraphics[width=0.5\textwidth]{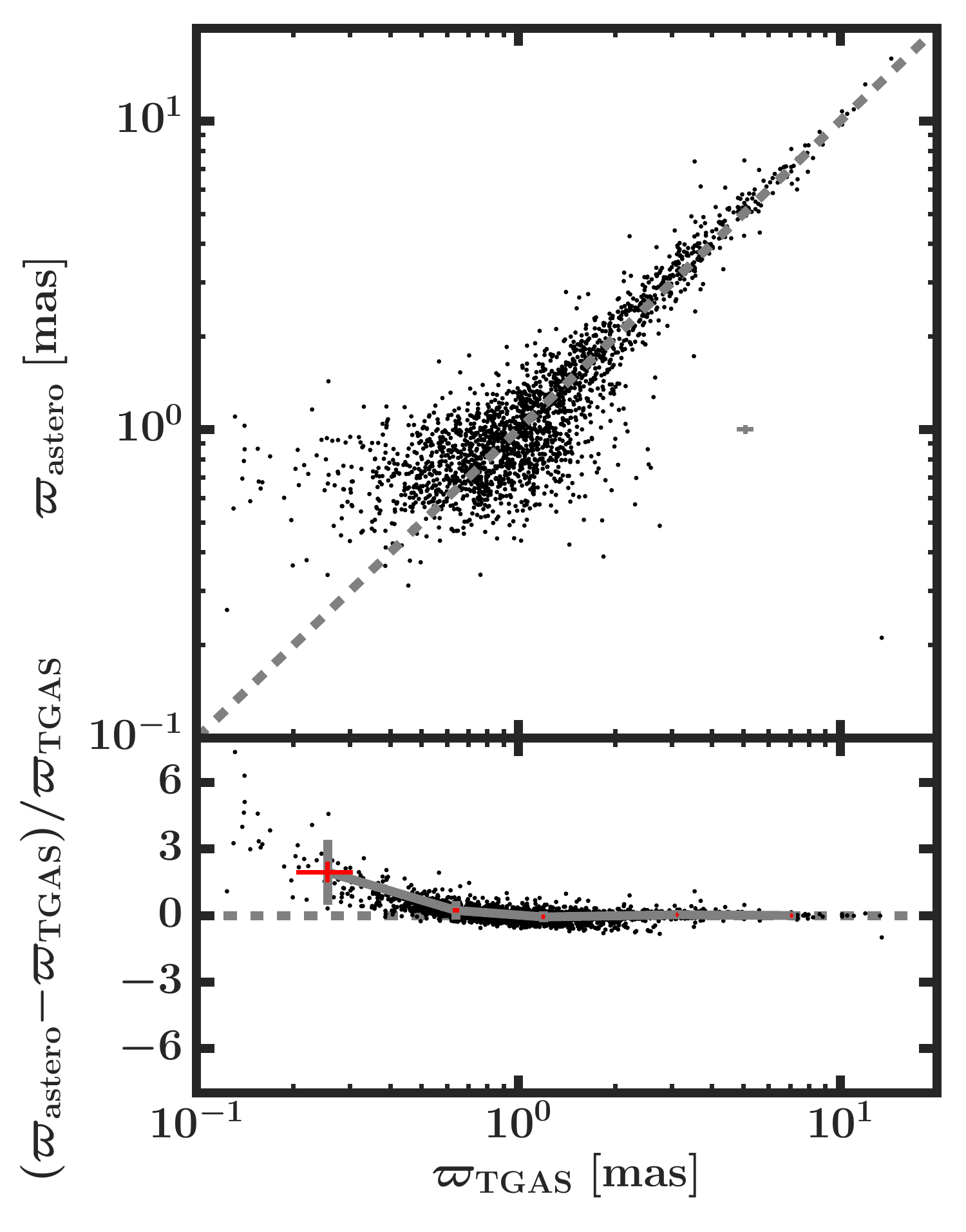}
}
\subfloat{
\centering
\includegraphics[width=0.5\textwidth]{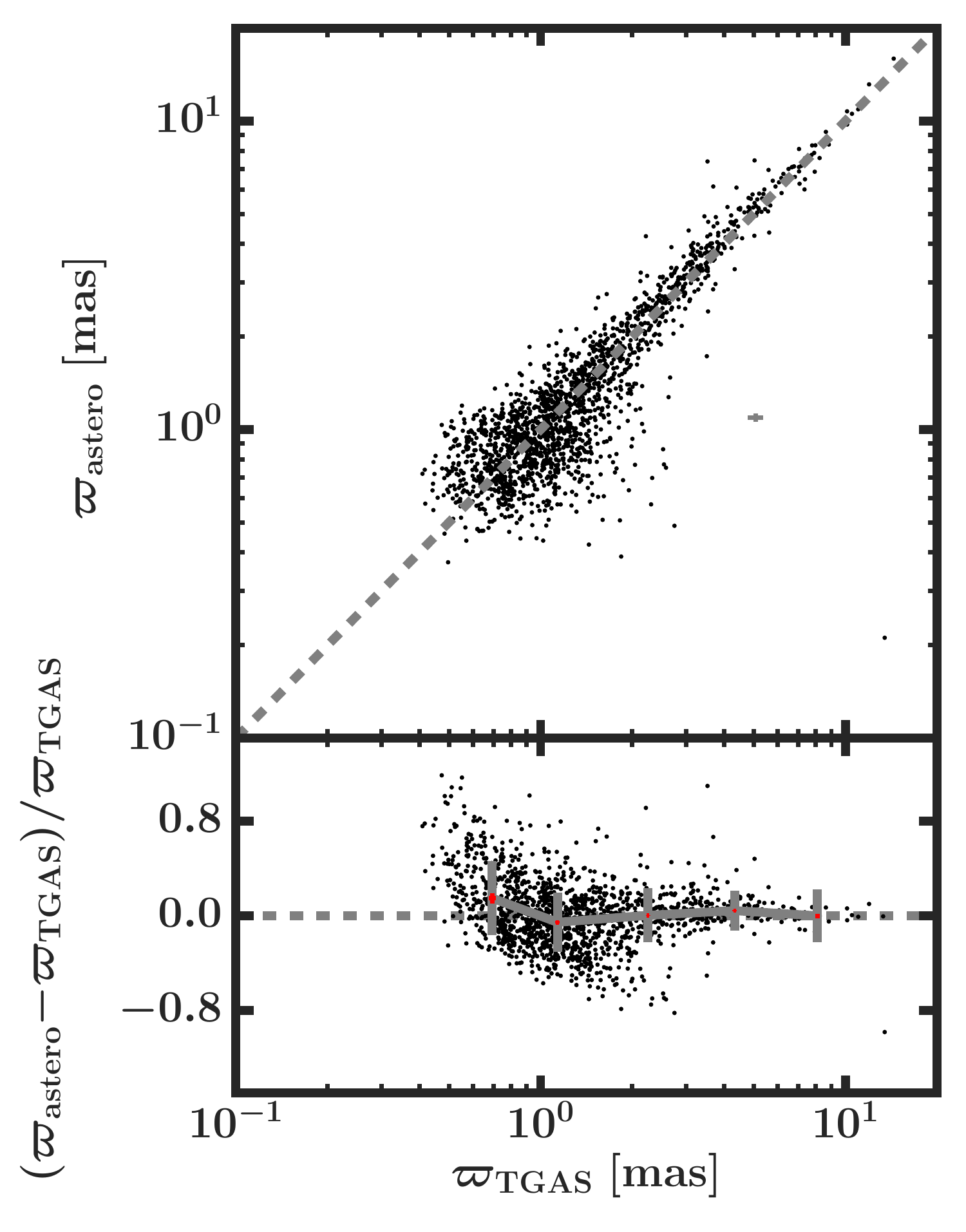}
}
\caption{Comparison of TGAS and asteroseismic parallax scales before
  ({ left})
  and after ({ right}) a signal-to-noise ratio cut of $\mathrm{SNR} >
  1.6$. { Median uncertainties are shown as the grey error bars in
    the top axes. Grey dashed lines show one-to-one relations. The
bottom axes show running medians of the fractional parallax differences (grey curves), with grey error bars representing the standard
deviation of the difference within each bin and red error bars
representing the statistical error on the median within each bin.}} 
\label{fig:plx_v_plx}
\end{figure*}

\subsection{Final TGAS-APOKASC sample}
\label{sec:sample}
\begin{figure}[htb!]
\centering
\includegraphics[width=0.5\textwidth]{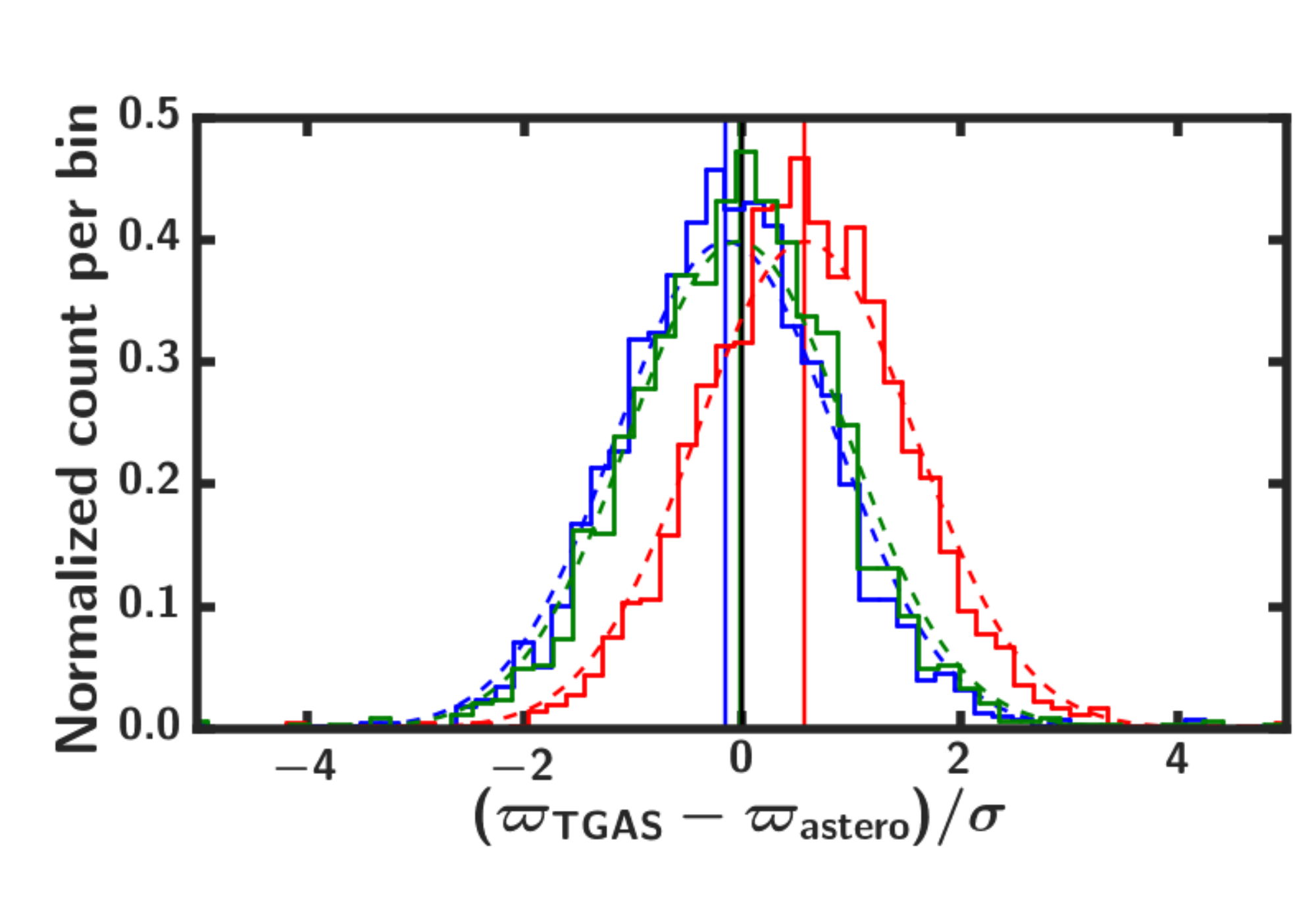}
\caption{The distribution of TGAS and asteroseismic parallax
  differences, in the sense of TGAS - asteroseismic, normalized by the
  statistical error on the difference, assuming normality. In blue is the
  { base sample of 1592 stars before a signal-to-noise ratio cut}. In
  red, a zeropoint correction is applied { to the base sample}, as recommended in
  \protect\cite{stassun&torres2016a}. In green { is the final
    TGAS-APOKASC sample of 1392 stars, which differs from the base
    sample by }a signal-to-noise cut
  of $\mathrm{SNR} > 1.6$ such that the median of the resulting distribution is
  equal to zero, to within the error on the median (a parallax
  difference of zero is shown as
  black vertical line).} 
\label{fig:zp}
\end{figure}

In Figure~\ref{fig:plx_v_plx}, we show a direct star-by-star
comparison of the two parallax scales. It is evident that at smaller
$\varpi$, there is a systematic offset between the two scales. This offset is expected
from comparing a precise asteroseismic { parallax} sample to a much
less precise sample of TGAS parallaxes: the large fractional errors on
TGAS parallax will tend to scatter to low parallax. We can
mitigate the offset by applying a signal-to-noise cut such that the
median of the difference between the two parallaxes is zero, to within
the error on the median. We show the original distribution and the
distribution of the parallax difference after a signal-to-noise cut of
$\mathrm{SNR} > 1.6$ in Figure~\ref{fig:zp}. A potential zeropoint offset in asteroseismic and TGAS
parallax scales is discussed in \S\ref{sec:zp_corr} (see also Huber et al.,
in press).
We use the high
signal-to-noise sample for the rest of the analysis
(TGAS-APOKASC sample), which
numbers 1392. %STAT

\section{Results}
\label{sec:results}
\subsection{Spatially-correlated offsets in asteroseismic and TGAS parallaxes}
\label{sec:results1}
\begin{figure}[htb!]
\centering
\includegraphics[width=0.5\textwidth]{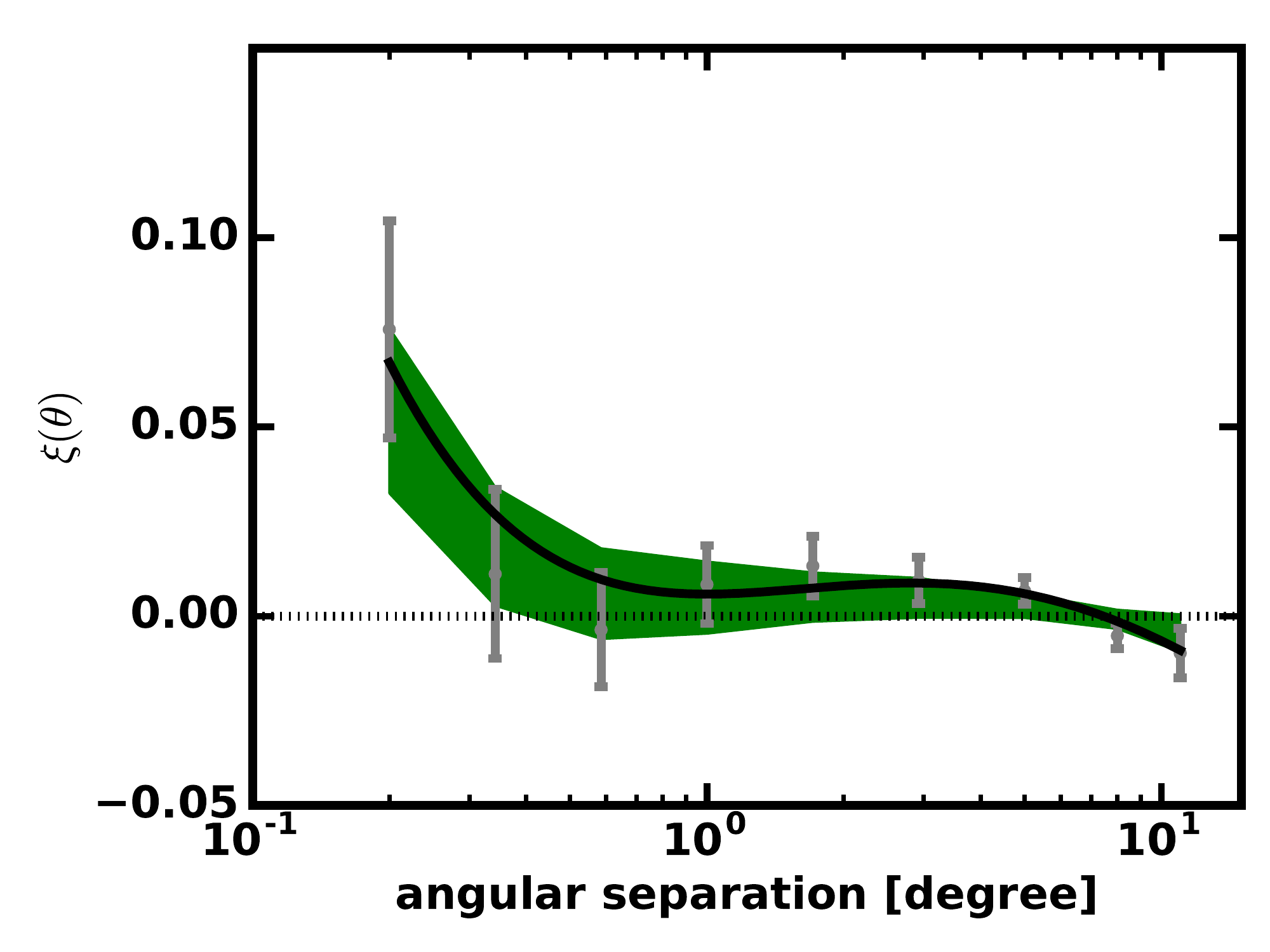}
\caption{The Pearson correlation coefficient, in bins of angular
  separation on the sky, which will be positive for correlated
  TGAS-asteroseismic parallax offsets, zero in the absence of correlated offsets, and
  negative for anti-correlated offsets. Error bars
  include a bootstrap and `systematic' error; the black line indicates
  the best-fitting model of the form in Equation~\ref{eq:model2}; the
  green band indicates the $68\%$ confidence interval of the recovered binned Pearson correlation coefficient, as
  computed from a mock catalogue of TGAS parallaxes assuming
  asteroseismic parallaxes as the true value and spatial correlations
  according to the black line. Refer to
  \S\ref{sec:results2} for details.}
\label{fig:ang_corr}
\end{figure}

Figure~\ref{fig:ang_corr} shows our main result, which is a measure
of the spatial correlation of the difference in the TGAS and
asteroseismic parallax scales for all 1392 giants in the TGAS-APOKASC % STAT
sample. Below, we first discuss the choice of a binned Pearson correlation
coefficient as a metric for the
spatially-correlated parallax difference and how it is calculated,
and then present model fits to the observed signal.
\subsection{Quantification using a binned Pearson correlation
  coefficient}
\label{sec:results2}
Our measure of the parallax offset spatial correlation is a binned Pearson correlation coefficient,
\begin{equation}
\xi (\theta) \equiv \frac{\sum_{i\neq j} (\plxtgasi - \plxasteroi)_{\theta}(\plxtgasj
  -
  \plxasteroj)_{\theta}}{\sqrt{((\plxtgasi - \plxasteroi)^2)_{\theta}((\plxtgasj-\plxasteroj)^2)_{\theta}}},
\label{eq:pearson}
\end{equation}
where $i$ and $j$ denote stars that are
separated by an angle, $\theta'$, such that $\theta - \Delta\theta/2 < \theta' <
\theta + \Delta\theta/2$ for a given angular bin size
$\Delta\theta$. Equation~\ref{eq:pearson} describes a Pearson correlation
coefficient computed in bins of angular separation. A value of $-1$
indicates that $\plxtgas$ and $\plxastero$ are perfectly
anti-correlated at that separation; a value of $+1$ indicates that
they are perfectly correlated; a value of $0$ indicates they are not
correlated. In the absence of spatial errors, then, we would expect a
null signal of zero at all angular separations. For errors that
increase at small separations, there would be a rise in $\xi(\theta)$
with decreasing $\theta$. A zeropoint error would result in a flat,
positive $\xi(\theta)$ for all $\theta$.

We compute error bars in the binned Pearson correlation coefficient
via bootstrapping \citep{loh2008}. Briefly, the sample of objects were divided into $N$
spatial regions, which were then sampled with replacement (meaning the
same region could be used multiple times in a single bootstrap sample)
$N$ times to
create a bootstrap sample; we divided the sample into
spatial regions according to their {\it Kepler} module. For $B$ such samples,
the { $100(1 - \alpha)\%$} confidence
intervals on each point in the binned Pearson correlation coefficient
were computed according to the bootstrap confidence interval \citep{davison&hinkley1997}:
\[
\left[2\hat{K} - K_{(B+1)(1 - \alpha/2)}, 2\hat{K} - K_{(B+1)\alpha/2}\right],
\]
where $K_{\mathrm{A}}$ is the A$^{th}$-ranked statistic (the binned Pearson correlation coefficient,
$K=\xi$, in this case) computed from a bootstrap sample and
$\hat{K}$ is the statistic computed using all the data.

In order to better characterize the errors on the statistic, we create
mock TGAS stellar catalogues, whose positions are the same as those in
the data, but whose parallaxes are drawn from the asteroseismic
parallaxes, and injected with spatial correlations
according to { Equation~\ref{eq:model2}, assuming Gaussian
statistics, with best-fitting
values from Table~\ref{tab:plx} (see \S\ref{sec:fitting})}. These fake TGAS parallaxes are
then used to compute the binned Pearson correlation coefficient
according to Equation~\ref{eq:pearson}, and the resulting distribution
of values at each angular bin are used to compute a $68\%$ confidence
interval --- a `systematic' error --- for the statistic. Note that our mock catalogue generation
assumes Gaussianity in the distribution of TGAS-asteroseismic parallax
difference, which we think is reasonable given the evident Gaussianity
of the distribution shown in Figure~\ref{fig:zp}. This `systematic'
error is shown as a green band in Figure~\ref{fig:ang_corr}.

We also provide alternate representations of the TGAS-asteroseismic
parallax offset in Figures~\ref{fig:ang_corr_treecorr} \&~\ref{fig:abs}. Both alternate
representations indicate a spatial dependence in the offset, in
agreement with the binned Pearson correlation coefficient. See the
Appendix for details on how these alternate measures are computed.

\begin{figure*}
\centering
\subfloat{
\includegraphics[width=0.8\textwidth]{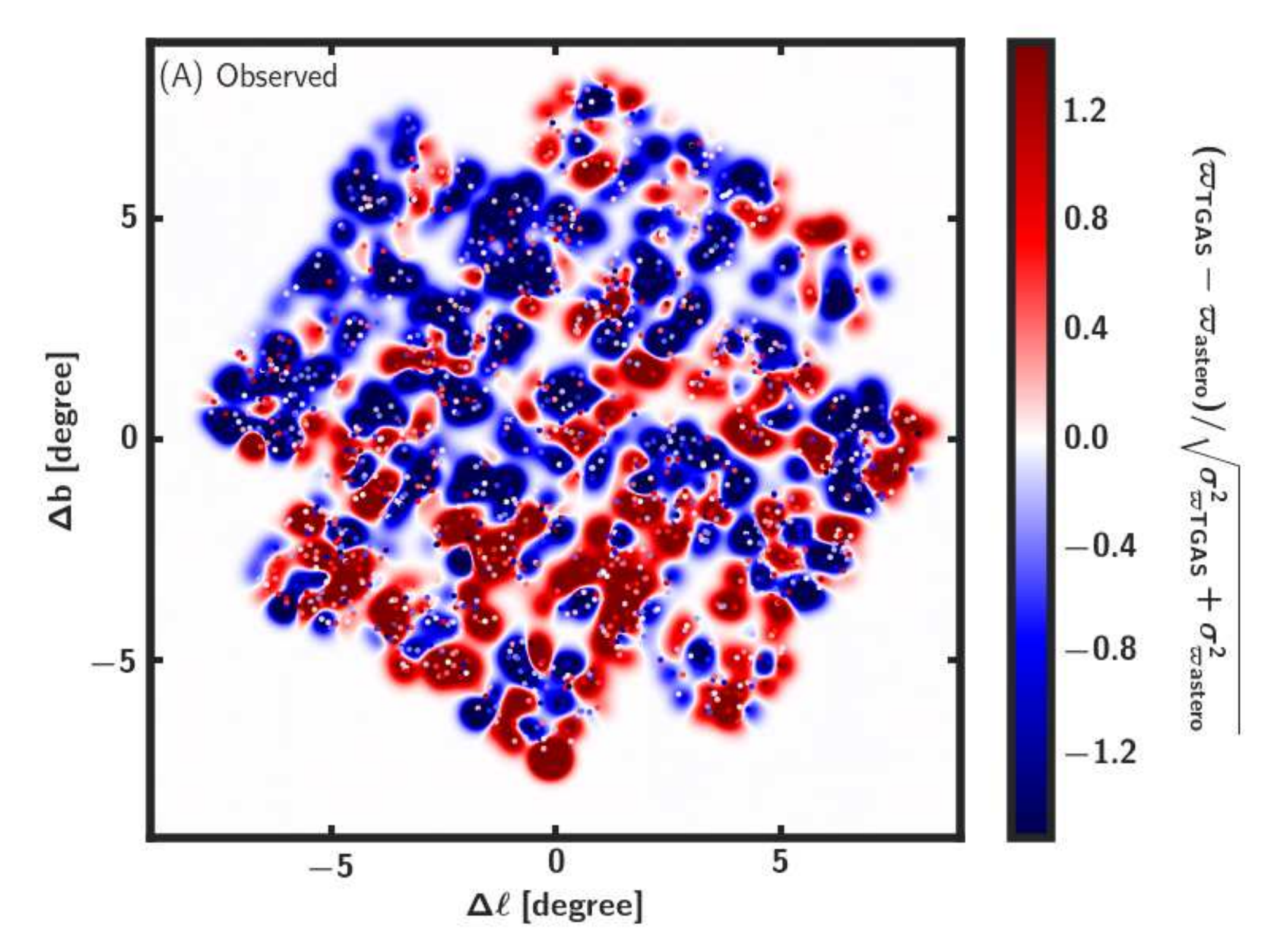}
}
\\
\subfloat{
\centering
\includegraphics[width=0.5\textwidth]{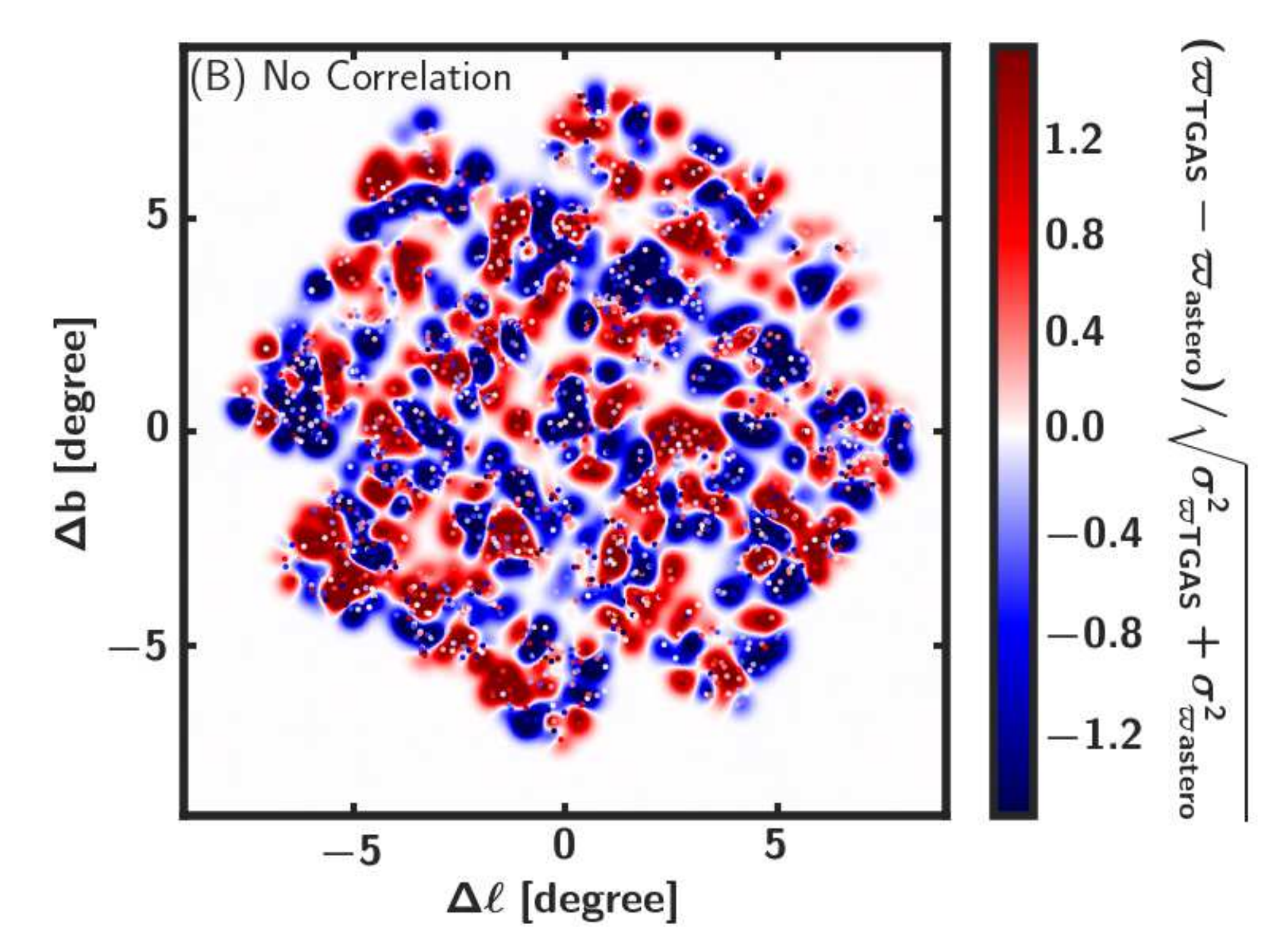}
}
\subfloat{
\centering
\includegraphics[width=0.5\textwidth]{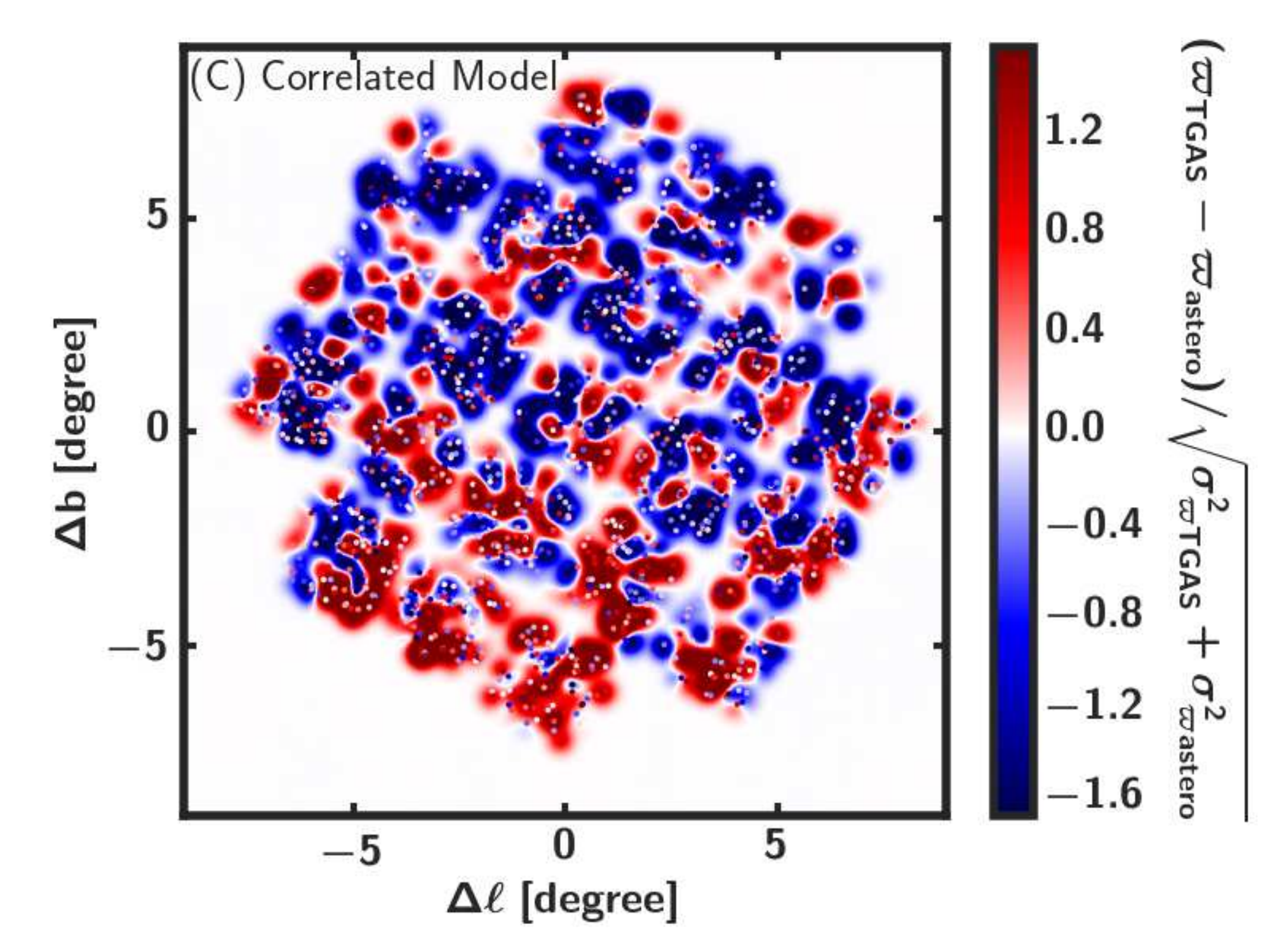}
}
\caption{The distribution of TGAS and asteroseismic parallax
  differences, in the sense of TGAS-asteroseismic, normalized by the
  statistical error on the difference, $\sigma \equiv
  \sqrt{\sigma_{\plxtgas}^2 + \sigma_{\plxastero}^2}$, as a function of
  position on the sky. The smoothed field is calculated by convolving
  the observed data with a Gaussian filter of standard deviation
  $0.2$deg. Panel a shows the signal in the TGAS-APOKASC
  sample; panel b shows a model with no spatial correlation; panel c
  shows the best-fitting model of panel a, with random spatial phase
  (see \S\ref{sec:fitting}).}
\label{fig:zoomin}
\end{figure*}

For visualization purposes, a spatial map of the TGAS-asteroseismic
parallax offset is plotted { in} Figure~\ref{fig:zoomin}a. Each point represents a star in the TGAS-APOKASC
sample and is colored by $(\plxtgas - \plxastero)/\sigma$, where
$\sigma$ is the quadrature sum of the statistical errors from $\plxtgas$ and $\plxastero$. A smoothed version
of the data is calculated by convolving these values by a Gaussian with
$0.2$deg standard deviation. For comparison, a map of parallax offset with no spatial
correlation is simulated in { Figure~\ref{fig:zoomin}}b. Whereas there is visible
structure shown in the observed TGAS-asteroseismic parallax
differences of { Figure~\ref{fig:zoomin}}a, there are no such correlated hot or cold spots
in Figure~\ref{fig:zoomin}b. For comparison, a map injected with spatial correlations according to
Equation~\ref{eq:model2}, assuming Gaussian statistics, with best-fitting values from
Table~\ref{tab:plx}, is shown in { Figure~\ref{fig:zoomin}}c. The spatial correlation
model of { Figure~\ref{fig:zoomin}}c qualitatively reproduces the patchwork structure seen in the data
({ Figure~\ref{fig:zoomin}}a).

\subsection{Fitting models to the observed spatially-correlated
  offset}
\label{sec:fitting}
We fit an analytic form to the observed spatial correlation in order
to determine a characteristic scale at which the systematic error is
important. We consider two models --- one with exponential spatial scale
dependence, as might be expected from the {\it Gaia} scanning strategy, in which characteristic spatial scales could be strongly imprinted in
the data.
The first model we fit to the binned Pearson correlation coefficient
is a purely exponential model of the form
\begin{equation}
\xi (\theta) = \rho_{\mathrm{max}} \exp{\left[-\theta /\theta_{1/2} \ln 2 \right]} + C.
\label{eq:model1}
\end{equation}

Note that in the above expression, $\rho_{\mathrm{max}}$ is an overall amplitude to
the spatially-varying component of $\xi$, and that $\theta_{1/2}$ represents the
angular scale at which spatial correlations are half of what they are
at the smallest scales.

We also fit a polynomial of the form
\begin{equation}
\xi (\theta) = A + B\log\theta + C(\log\theta)^2 + D(\log\theta)^3 + E(\log\theta)^4.
\label{eq:model2}
\end{equation}

We computed best-fitting parameters and their associated uncertainties by
fitting with the \texttt{PYTHON} MCMC routine of \texttt{emcee}
\citep{foreman-mackey+2013} with
a covariance matrix calculated from the bootstrap sample, whose
diagonal is added in quadrature with the `systematic' error, described above.

In Table~\ref{tab:plx}, we provide the resulting best-fitting
parameters for the spatially-correlated parallax offset models. We
also provide best-fitting parameters for
the RGB and RC sub-samples separately (see \S\ref{sec:rgb_rc}). According to the
Akaike Information Criterion \citep{akaike1973}, $AIC \equiv 2k - \ln
\mathcal{L}$, where $k$ is the number of degrees of freedom and
$\ln \mathcal{L} = -0.5 \chi^2$ is the log-likelihood, Equation~\ref{eq:model1} is preferred
in the fits to binned Pearson correlation coefficient for RGBs { and} RCs,
but Equation~\ref{eq:model2} is preferred for the combined TGAS-APOKASC sample. We take preference of model 1 over model 2 to be
$AIC_{\mathrm{model 1}} - AIC_{\mathrm{model 2}} < -2$. We refer to
the best-fitting model as Equation~\ref{eq:model2}, and recommend the fits to this polynomial model
for characterizing covariance matrices. For completeness, we have also
compared preference for both models to
a null model of zero at all angular scales, finding that the null
model is never preferred.

Systematic offsets at a given angular scale, $\theta$,
$\sigma_{\mathrm{sys}}(\theta)$, are also reported in
Table~\ref{tab:plx}. They are calculated according to
\begin{equation}
\sigma_{\mathrm{sys}}(\theta) = \sqrt{|\xi(\theta)|\sigma^2},
\label{eq:cov}
\end{equation}
where $\sigma \equiv \sqrt{\sigma_{\plxtgas}^2 +
    \sigma_{\plxastero}^2}$. Confidence intervals on $\sigma_{\mathrm{sys}}$ are
computed using a covariance matrix built from the MCMC chains from
model-fitting (see above), according to which a representative
distribution of model parameters is drawn, and for which a resulting
distribution of possible $\sigma_{\mathrm{sys}}$ is computed. Note that the sense of this systematic
offset between the two parallax scales is not indicated, as it will
vary as a function of absolute position on the sky. One can see, for instance,
the regions where the sign change of the offset switches in
Figure~\ref{fig:zoomin}. Rather, signs on this
systematic offset provided in Table~\ref{tab:plx} indicate correlation (positive)
versus anti-correlation (negative). 

Our best-fitting polynomial model using the entire
TGAS-APOKASC sample yields
parallax offsets at the smallest separations of $0.059^{+0.004}_{-0.004}$mas and
$0.011^{+0.006}_{-0.004}$mas at % STAT
spacial scales of $\theta \approx 0.3$deg and $\theta \approx
8$deg, respectively. %STAT

\floattable
\begin{deluxetable}{ccccccccccccc}
%  \tablecolumns{11} 
%  \tablewidth{9in}
  \tabletypesize{\scriptsize}
  \rotate
  \tablehead{\colhead{sample} & \colhead{model} &
    \colhead{$\rho_{\mathrm{max}} / A$} & \colhead{$\theta_{1/2} / B$} &
    \colhead{$C$} & \colhead{$D$} & \colhead{$E$} & \colhead{$\chi^2/dof$} & \colhead{$\sigma_{\mathrm{{sys}}}(\theta = \theta_{1/2})$}
    & \colhead{$\sigma_{\mathrm{{sys}}}(\theta = 0.0\deg)$}&
    \colhead{$\sigma_{\mathrm{{sys}}}(\theta = 0.3\deg)$} &
    \colhead{$\sigma_{\mathrm{{sys}}}(\theta = 1.0\deg)$} &
    \colhead{$\sigma_{\mathrm{{sys}}}(\theta = 8.0\deg)$}}
  \tablecaption{Fits to models of the spatially-correlated TGAS-asteroseismic parallax offset  \label{tab:plx}}
  \startdata
ALL & $Eq.\ref{eq:model1}^{*}$ & $0.059^{+0.045}_{-0.045}$ &
$0.110^{+0.040 \circ}_{-0.040}$ & $0.0032^{+0.0005}_{-0.0005}$ & & &
7.287 & $0.050^{+0.016}_{-0.022}$mas & $0.078^{+0.025}_{-0.037}$mas &
$0.030^{+0.010}_{-0.011}$mas & $0.018^{+0.001}_{-0.002}$mas &
$0.017^{+0.001}_{-0.001}$mas \\
ALL & $Eq.\ref{eq:model2}^{*+}$ & $0.006^{+0.002}_{-0.002}$ &
$-0.022^{+0.575 \circ}_{-0.575}$ & $0.009^{+0.001}_{-0.001}$
&$-0.007^{+0.001}_{-0.001}$ & $0.0010^{+0.0003}_{-0.0003}$ & 4.850 &
\nodata & \nodata &
$0.059^{+0.004}_{-0.004}$mas & $0.024^{+0.004}_{-0.005}$mas &
$-0.011^{+0.006}_{-0.004}$mas \\
RC & $Eq.\ref{eq:model1}^{*}$ & $0.085^{+0.019}_{-0.019}$ &
$4.970^{+2.035 \circ}_{-2.035}$ & $-0.0310^{+0.0172}_{-0.0172}$ & & &
0.976 & $0.031^{+0.006}_{-0.009}$mas & $0.071^{+0.006}_{-0.007}$mas &
$0.068^{+0.007}_{-0.007}$mas & $0.062^{+0.006}_{-0.006}$mas &
$-0.017^{+0.031}_{-0.011}$mas \\
RC & $Eq.\ref{eq:model2}^{*}$ & $0.045^{+0.008}_{-0.008}$ &
$-0.025^{+0.583 \circ}_{-0.583}$ & $-0.014^{+0.006}_{-0.006}$
&$-0.005^{+0.004}_{-0.004}$ & $0.0028^{+0.0011}_{-0.0011}$ & 1.215 &
\nodata & \nodata &
$0.058^{+0.011}_{-0.014}$mas & $0.065^{+0.005}_{-0.006}$mas &
$-0.025^{+0.011}_{-0.006}$mas \\
RGB & $Eq.\ref{eq:model1}^{*}$ & $0.104^{+0.027}_{-0.027}$ &
$7.590^{+1.793 \circ}_{-1.793}$ & $-0.0367^{+0.0206}_{-0.0206}$ & & &
0.230 & $0.037^{+0.009}_{-0.013}$mas & $0.078^{+0.009}_{-0.009}$mas &
$0.078^{+0.008}_{-0.010}$mas & $0.073^{+0.009}_{-0.010}$mas &
$0.035^{+0.010}_{-0.015}$mas \\
RGB & $Eq.\ref{eq:model2}^{*}$ & $0.072^{+0.018}_{-0.018}$ &
$-0.036^{+0.576 \circ}_{-0.576}$ & $-0.005^{+0.004}_{-0.004}$
&$-0.009^{+0.004}_{-0.004}$ & $0.0027^{+0.0014}_{-0.0014}$ & 0.044 &
\nodata & \nodata &
$0.091^{+0.013}_{-0.015}$mas & $0.082^{+0.010}_{-0.011}$mas &
$0.043^{+0.011}_{-0.014}$mas \\
\enddata
\tablecomments{Best-fitting parameters for Equations~\protect\ref{eq:model1}
  \&~\protect\ref{eq:model2} for the spatially-correlated
  TGAS-asteroseismic parallax offset, with $68\%$
  confidence interval errors. A positive (negative) value in the last five columns
  indicates the systematic offset is a positive correlation (an
  anti-correlation). An asterisk indicates preference of the model
  over a null signal based on the AIC criterion (see text). A plus
  sign indicates that the model of Equation~\protect\ref{eq:model2} is
  preferred over the model of Equation~\protect\ref{eq:model1} according to the AIC criterion.}
\end{deluxetable}

\floattable
\begin{deluxetable}{ccccccccccc}
  \tabletypesize{\small}
  \rotate
  \tablehead{\colhead{observable} & \colhead{model} & \colhead{$\rho_{\mathrm{max}}$} & \colhead{$\theta_{1/2}$} & \colhead{$C$} & \colhead{$\chi^2/dof$} & \colhead{$\sigma_{\mathrm{{sys}}}(\theta = \theta_{1/2})$}
  & \colhead{$\sigma_{\mathrm{{sys}}}(\theta = 0.0\deg)$}&
    \colhead{$\sigma_{\mathrm{{sys}}}(\theta = 0.3\deg)$} &
    \colhead{$\sigma_{\mathrm{{sys}}}(\theta = 1.0\deg)$} &
    \colhead{$\sigma_{\mathrm{{sys}}}(\theta = 8.0\deg)$}}
  \tablecaption{Spatial correlation in observables   \label{tab:obs}}
  \startdata
$\nu_{\mathrm{max}}$ & $Eq.\ref{eq:model1}^{*}$ &
  $-0.004^{+0.010}_{-0.010}$ &  $6.085^{+2.628 \circ}_{-2.628}$ &
  $0.0011^{+0.0011}_{-0.0011}$ & 4.160 &
  $-0.026^{+0.081}_{-0.049}$$\mu$Hz &
  $-0.046^{+0.126}_{-0.057}$$\mu$Hz &
  $-0.050^{+0.131}_{-0.053}$$\mu$Hz &
  $-0.039^{+0.115}_{-0.056}$$\mu$Hz &
  $-0.022^{+0.072}_{-0.042}$$\mu$Hz \\
$\Delta\nu$ & $Eq.\ref{eq:model1}^{*}$ & $-0.0036^{+0.0027}_{-0.0027}$
  &  $5.8721^{+2.7097 \circ}_{-2.7097}$ & $0.0009^{+0.0011}_{-0.0011}$
  & 4.6685 & $-0.0008^{+0.0012}_{-0.0005}$$\mu$Hz &
  $-0.0015^{+0.0010}_{-0.0006}$$\mu$Hz &
  $-0.0014^{+0.0010}_{-0.0005}$$\mu$Hz &
  $-0.0013^{+0.0009}_{-0.0005}$$\mu$Hz &
  $-0.0007^{+0.0013}_{-0.0004}$$\mu$Hz \\
$T_{\mathrm{eff, IRFM}}$ & $Eq.\ref{eq:model1}^{*}$ &
  $0.007^{+0.021}_{-0.021}$ &  $0.845^{+3.170 \circ}_{-3.170}$ &
  $-0.0011^{+0.0013}_{-0.0013}$ & 1.445 & $1.655^{+3.782}_{-5.467}$K &
  $2.358^{+3.013}_{-6.215}$K & $2.349^{+3.066}_{-5.968}$K &
  $1.787^{+4.162}_{-5.457}$K & $-0.815^{+15.744}_{-1.744}$K \\
$A_V$ & $Eq.\ref{eq:model1}^{*}$ & $0.923^{+0.057}_{-0.057}$ &
  $4.805^{+0.516 \circ}_{-0.516}$ & $-0.3844^{+0.0546}_{-0.0546}$ &
  0.920 & $0.022^{+0.001}_{-0.001}$mag & $0.059^{+0.001}_{-0.001}$mag
  & $0.057^{+0.001}_{-0.001}$mag & $0.051^{+0.001}_{-0.001}$mag &
  $-0.024^{+0.001}_{-0.001}$mag \\
$\feh$ & $Eq.\ref{eq:model1}^{*}$ & $0.025^{+0.005}_{-0.005}$ &
  $8.144^{+1.581 \circ}_{-1.581}$ & $-0.0152^{+0.0042}_{-0.0042}$ &
  1.529 & $-0.0014^{+0.0002}_{-0.0001}$ & $0.0025^{+0.0002}_{-0.0003}$
  & $0.0024^{+0.0002}_{-0.0003}$ & $0.0022^{+0.0002}_{-0.0002}$ &
  $-0.0013^{+0.0002}_{-0.0001}$ \\
$J$ & $Eq.\ref{eq:model1}^{*}$ & $0.002^{+0.014}_{-0.014}$ &
  $2.996^{+3.412 \circ}_{-3.412}$ & $-0.0006^{+0.0012}_{-0.0012}$ &
  5.029 & $-0.0003^{+0.0027}_{-0.0017}$mag &
  $0.0007^{+0.0017}_{-0.0031}$mag & $0.0006^{+0.0020}_{-0.0029}$mag &
  $0.0005^{+0.0021}_{-0.0026}$mag & $-0.0004^{+0.0034}_{-0.0010}$mag
  \\
$H$ & $Eq.\ref{eq:model1}^{*}$ & $0.008^{+0.017}_{-0.017}$ &
  $0.636^{+3.311 \circ}_{-3.311}$ & $-0.0009^{+0.0014}_{-0.0014}$ &
  2.299 & $0.0015^{+0.0020}_{-0.0034}$mag &
  $0.0017^{+0.0014}_{-0.0036}$mag & $0.0015^{+0.0018}_{-0.0037}$mag &
  $0.0014^{+0.0025}_{-0.0035}$mag & $-0.0001^{+0.0109}_{-0.0011}$mag
  \\
$K_{\mathrm{s}}$ & $Eq.\ref{eq:model1}^{}$ & $0.008^{+0.014}_{-0.014}$
  &  $2.480^{+3.351 \circ}_{-3.351}$ & $-0.0013^{+0.0020}_{-0.0020}$ &
  15.329 & $0.0007^{+0.0028}_{-0.0021}$mag &
  $0.0015^{+0.0010}_{-0.0030}$mag & $0.0014^{+0.0013}_{-0.0029}$mag &
  $0.0011^{+0.0014}_{-0.0025}$mag & $-0.0002^{+0.0066}_{-0.0010}$mag
  \\
$g$ & $Eq.\ref{eq:model1}^{*}$ & $-0.008^{+0.002}_{-0.002}$ &
      $3.049^{+2.163 \circ}_{-2.163}$ & $0.0017^{+0.0016}_{-0.0016}$ &
      4.763 & $-0.0005^{+0.0001}_{-0.0001}$mag &
      $-0.0008^{+0.0001}_{-0.0001}$mag &
      $-0.0007^{+0.0001}_{-0.0001}$mag &
      $-0.0007^{+0.0001}_{-0.0001}$mag &
      $0.0002^{+0.0001}_{-0.0003}$mag \\
$r$ & $Eq.\ref{eq:model1}^{*}$ & $-0.009^{+0.001}_{-0.001}$ &
      $4.409^{+2.377 \circ}_{-2.377}$ & $0.0029^{+0.0019}_{-0.0019}$ &
      4.857 & $-0.0003^{+0.0001}_{-0.0001}$mag &
      $-0.0007^{+0.0001}_{-0.0001}$mag &
      $-0.0007^{+0.0001}_{-0.0001}$mag &
      $-0.0006^{+0.0001}_{-0.0001}$mag &
      $0.0002^{+0.0001}_{-0.0001}$mag \\
  \enddata
  \tablecomments{Best-fitting parameters for Equation~\protect\ref{eq:model1}
  for spatial correlations in observables, with $68\%$
  confidence interval errors. A positive (negative) value in the last five columns
  indicates the systematic offset is a positive correlation (an
  anti-correlation). An asterisk indicates preference of the model
  over a null signal based on the AIC criterion (see text).}
\end{deluxetable}

\section{Discussion}
\label{sec:discussion}
\subsection{`Systematic' error in Pearson statistic}
\label{sec:model}
As discussed in \S\ref{sec:results}, we estimate `systematic' errors in the
binned Pearson correlation coefficient by creating mock TGAS
parallax catalogues. We do this because we expect bootstrap errors to
underestimate the true error in the Pearson statistic for at least three reasons: (1) bootstrap sampling cannot account for the finite spatial extent of the
{\it Kepler} field of view, which will affect the Pearson correlation
coefficient in the largest angular separation bins (those comparable
to the length of the side of the {\it Kepler} field of view); (2) by drawing
distributions of parallaxes with correlated error patterns of a
random spatial phase (where spatial phase determines the locations on
the sky of the hot and cold spots in Figure~\ref{fig:zoomin}), one marginalizes over the phase of the spatial
correlation in a way that cannot be done with the single DR1 TGAS
parallax catalogue; (3) bootstrap sampling of the Pearson correlation
coefficient does not take into account statistical errors in the
parallaxes like creating sets of mock TGAS parallax catalogues does.

By adding the
`systematic' and bootstrap errors in quadrature,
we have likely overestimated the errors on the binned Pearson correlation
coefficient, and so the significance of our result is conservative.

\subsection{Correlation as a function of evolutionary type}
\label{sec:rgb_rc}
\begin{figure}[htb!]
\centering
\includegraphics[width=0.5\textwidth]{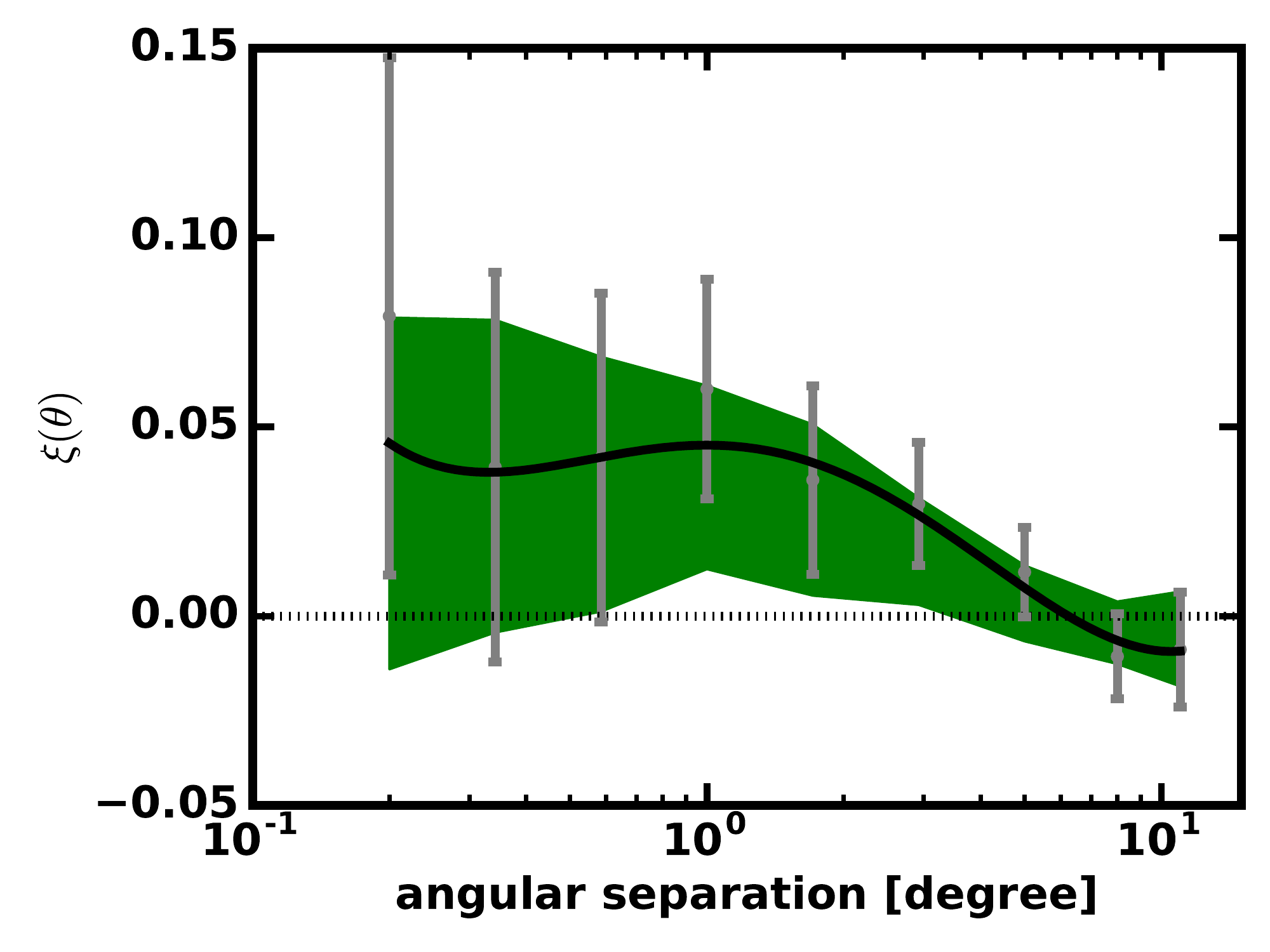}
\caption{Same as Figure~\ref{fig:ang_corr}, but calculated using only
  RCs.}
\label{fig:ang_corr_rc}
\end{figure}

\begin{figure}[htb!]
\centering
\includegraphics[width=0.5\textwidth]{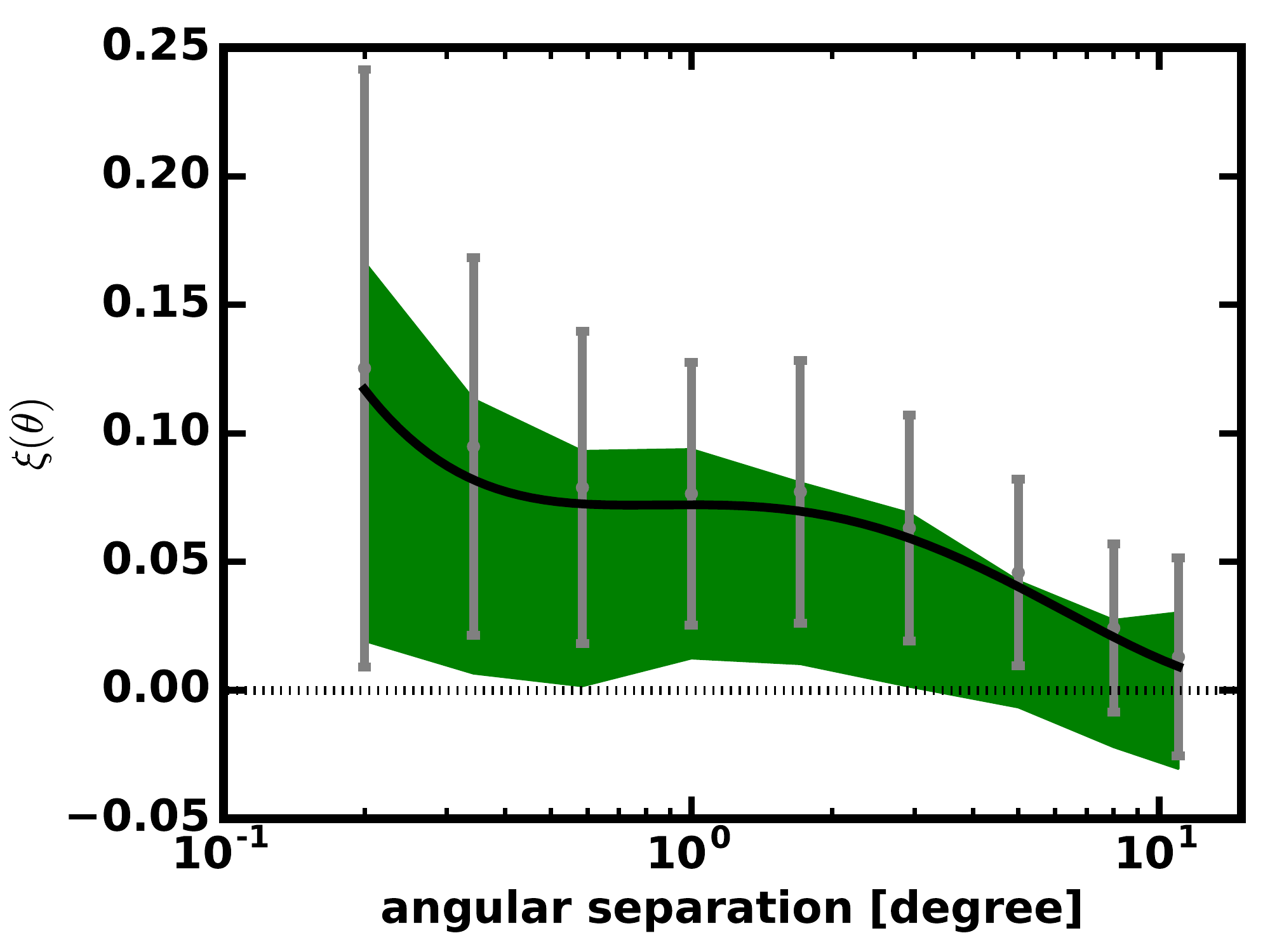}
\caption{Same as Figure~\ref{fig:ang_corr}, but calculated using only
  RGBs.}
\label{fig:ang_corr_rgb}
\end{figure}

As noted in \S\ref{sec:data}, there is evidence that RGBs and RCs
obey different asteroseismic scaling relations, which could lead to
systematic differences in their parallaxes, and perhaps a difference
in a spatially-correlated offsets from TGAS parallaxes. We present Figures~\ref{fig:ang_corr_rc} \&~\ref{fig:ang_corr_rgb} in
order to investigate whether the observed spatial correlations in
parallax offset vary with stellar type. We find the two samples to
yield consistent signals at all scales, with the RGB sample exhibiting
a mildly larger amplitude. Moreover, when averaged over the entire {\it Kepler}
field, there do not seem to be significant differences in the
TGAS and asteroseismic parallax scales as a function of evolutionary
type (see \S\ref{sec:zp_corr} and
Huber et al., in press). We note, furthermore, that this observation also suggests
that intrinsic spatial correlations of extinction are not contributing
significantly to the signal, because in that case we would expect RC
parallaxes to be more spatially-correlated than those of RGBs, since they have larger
distances on average than RGBs. However, the RC sample shows mildly
{\it smaller}, not larger, correlation coefficients than
the RGB sample. We discuss asteroseismic parallax scale systematics
further in \S\ref{sec:obs}.

\subsection{Bias in observables}
As we note in \S\ref{sec:rgb_rc}, spatial correlations should not arise from
global biases in the asteroseismic parallaxes. For completeness,
however, we perform several checks on the the reliability of the
quantities that are used to compute asteroseismic parallaxes.

\begin{figure}[htb!]
\includegraphics[width=0.5\textwidth]{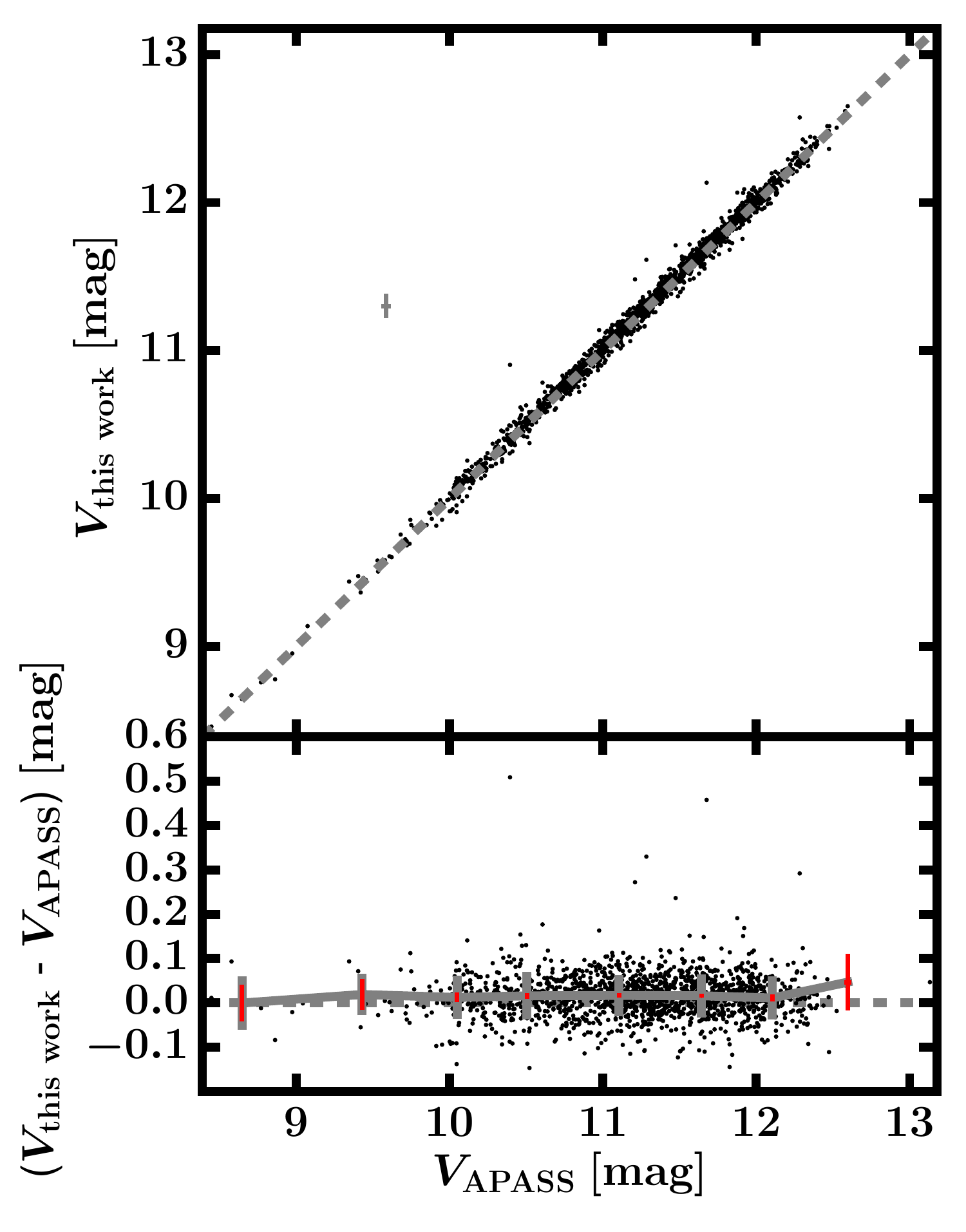}
\caption{$V$ magnitudes as derived in this work versus those from
  APASS \protect\citep{henden&munari2014}, which are consistent within
  statistical errors (median error bar is shown in the top panel). { Grey dashed lines show one-to-one relations. The
bottom axes show running medians of the fractional parallax
differences (grey curves), with grey error bars representing the
standard deviation of the difference within each bin and red error bars
representing the statistical error on the median within each bin.}}
\label{fig:v_apass}
\end{figure}

First, we confirmed that our re-reddened $V$ agree with those from
APASS \citep{henden&munari2014} to within statistical errors, as shown
in Figure~\ref{fig:v_apass}. 

We also tested the effect of using the APOGEE spectroscopic
temperature versus an IRFM temperature in computing asteroseismic
parallax. The temperature scale does not remove the observed spatial
correlation (see \S\ref{sec:obs}). Nevertheless, there is a different
zeropoint offset when using the IRFM temperature scale, which
originates from the IRFM being systematically hotter than the APOGEE
spectroscopic temperature (see Figure~\ref{fig:apogee_v_irfm}).

\begin{figure*}
\subfloat{
\centering
   \includegraphics[width=0.5\textwidth]{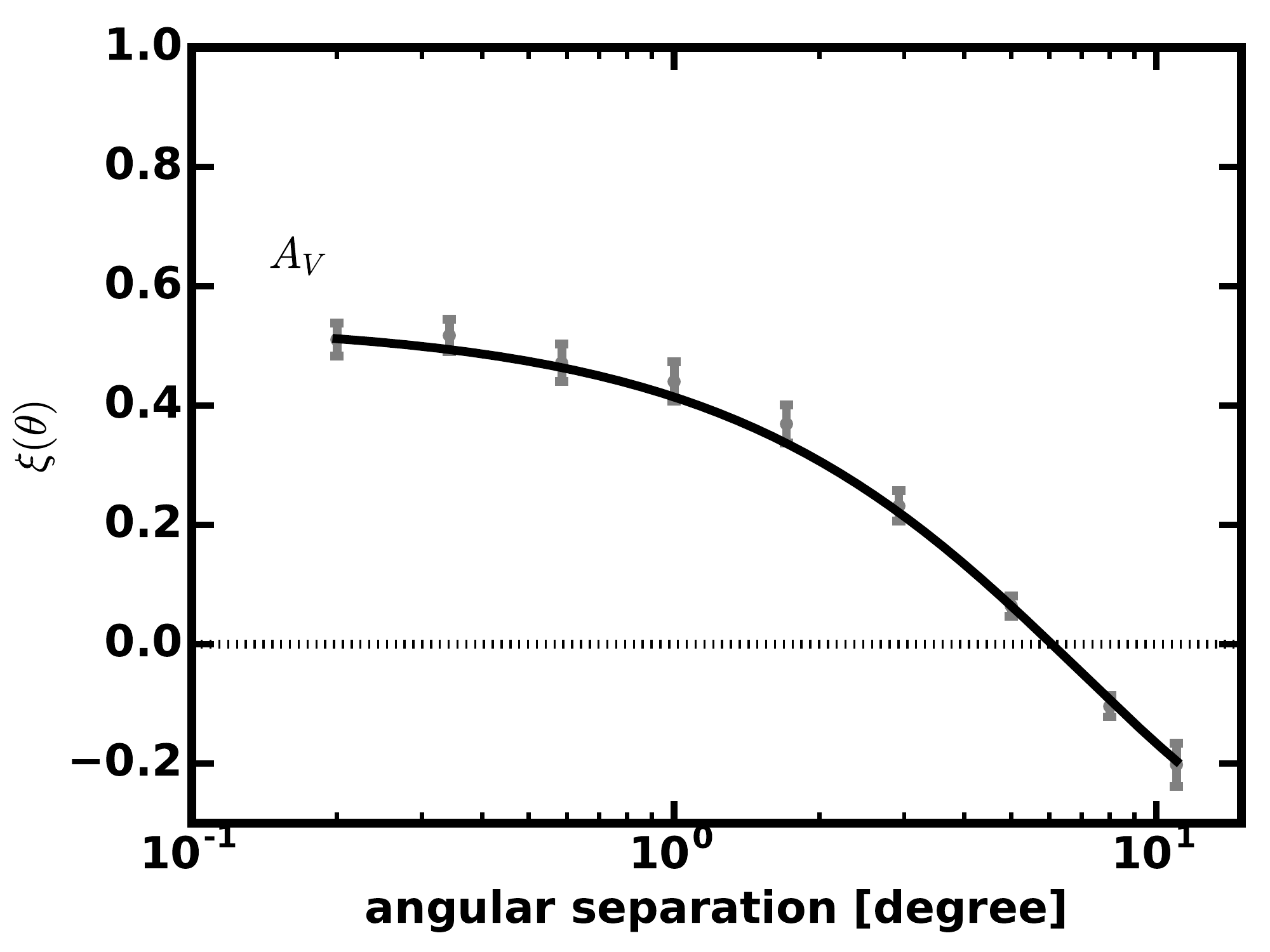}
}
\subfloat{
\centering
   \includegraphics[width=0.5\textwidth]{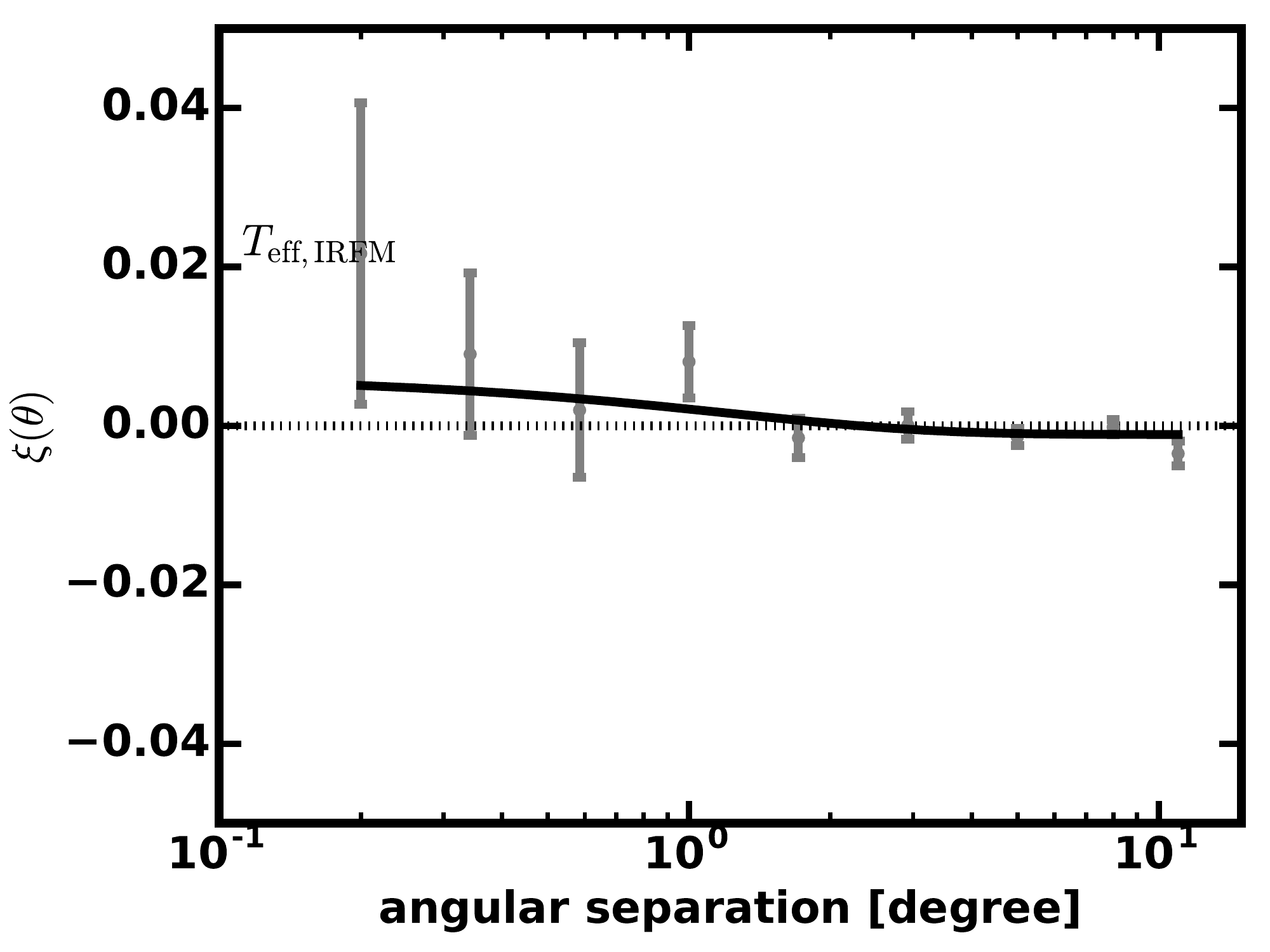}
}
\\
\subfloat{
\centering
   \includegraphics[width=0.5\textwidth]{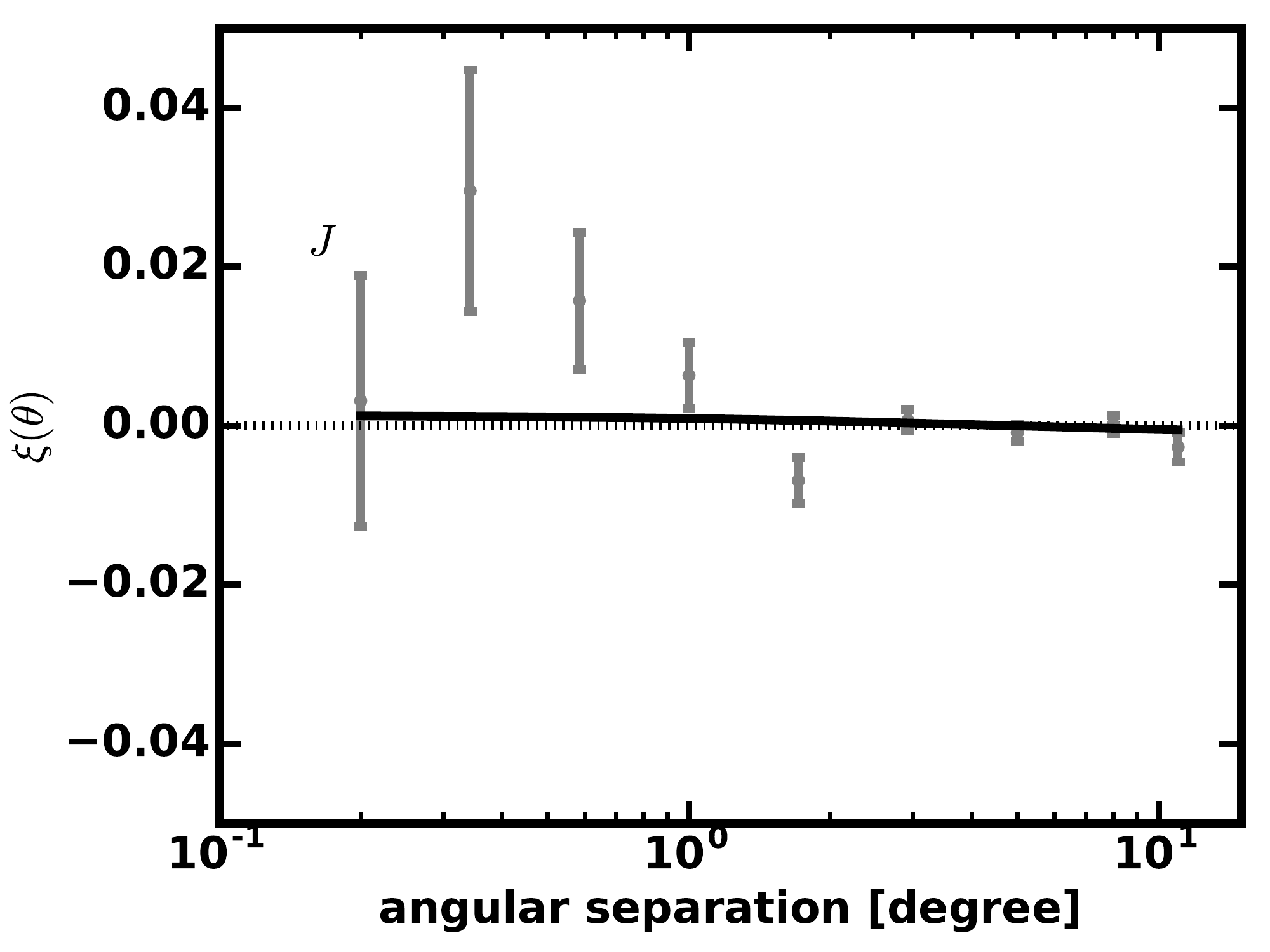}
}
\subfloat{
\centering
   \includegraphics[width=0.5\textwidth]{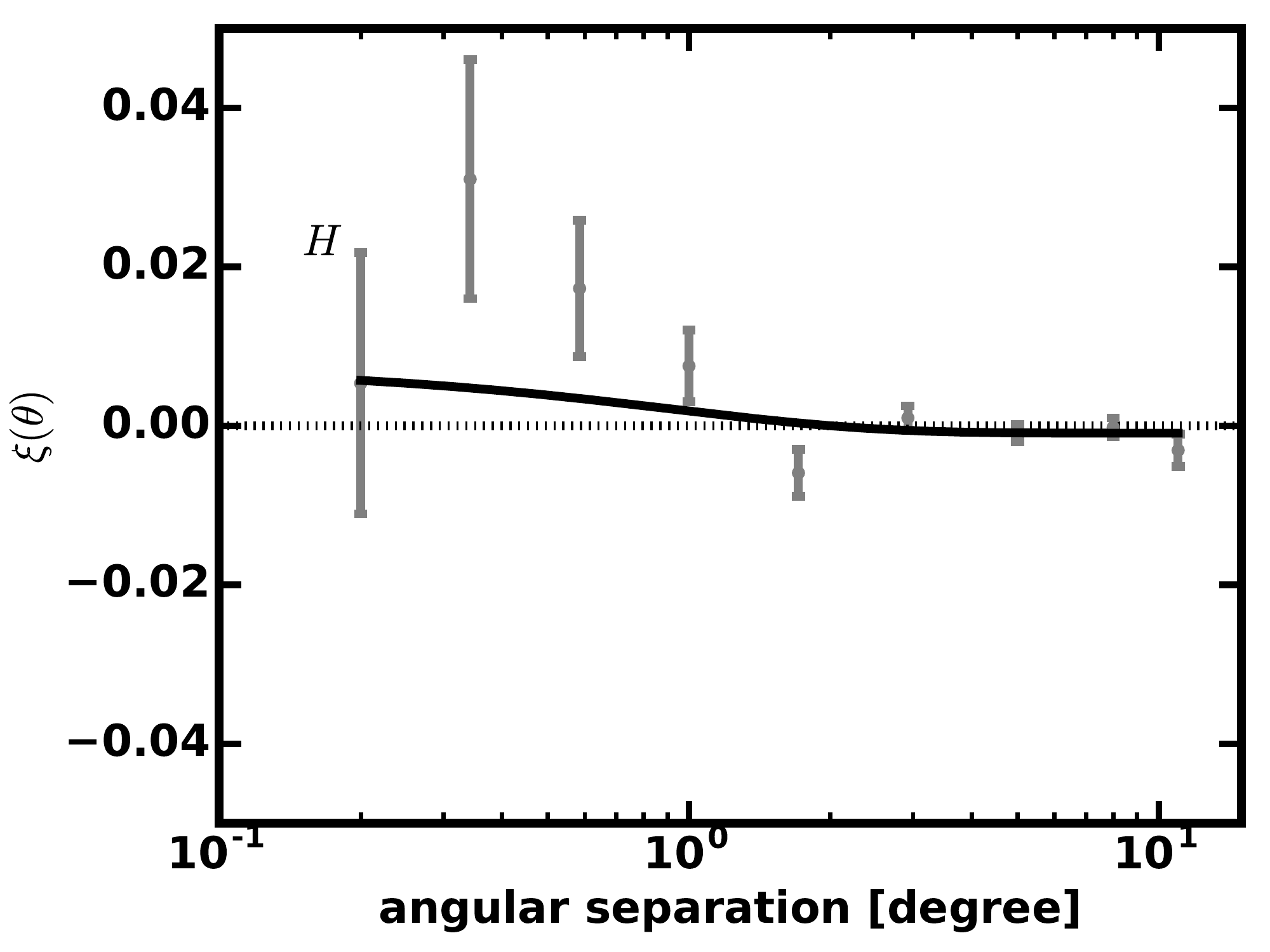}
}
\\
\subfloat{
\centering
   \includegraphics[width=0.5\textwidth]{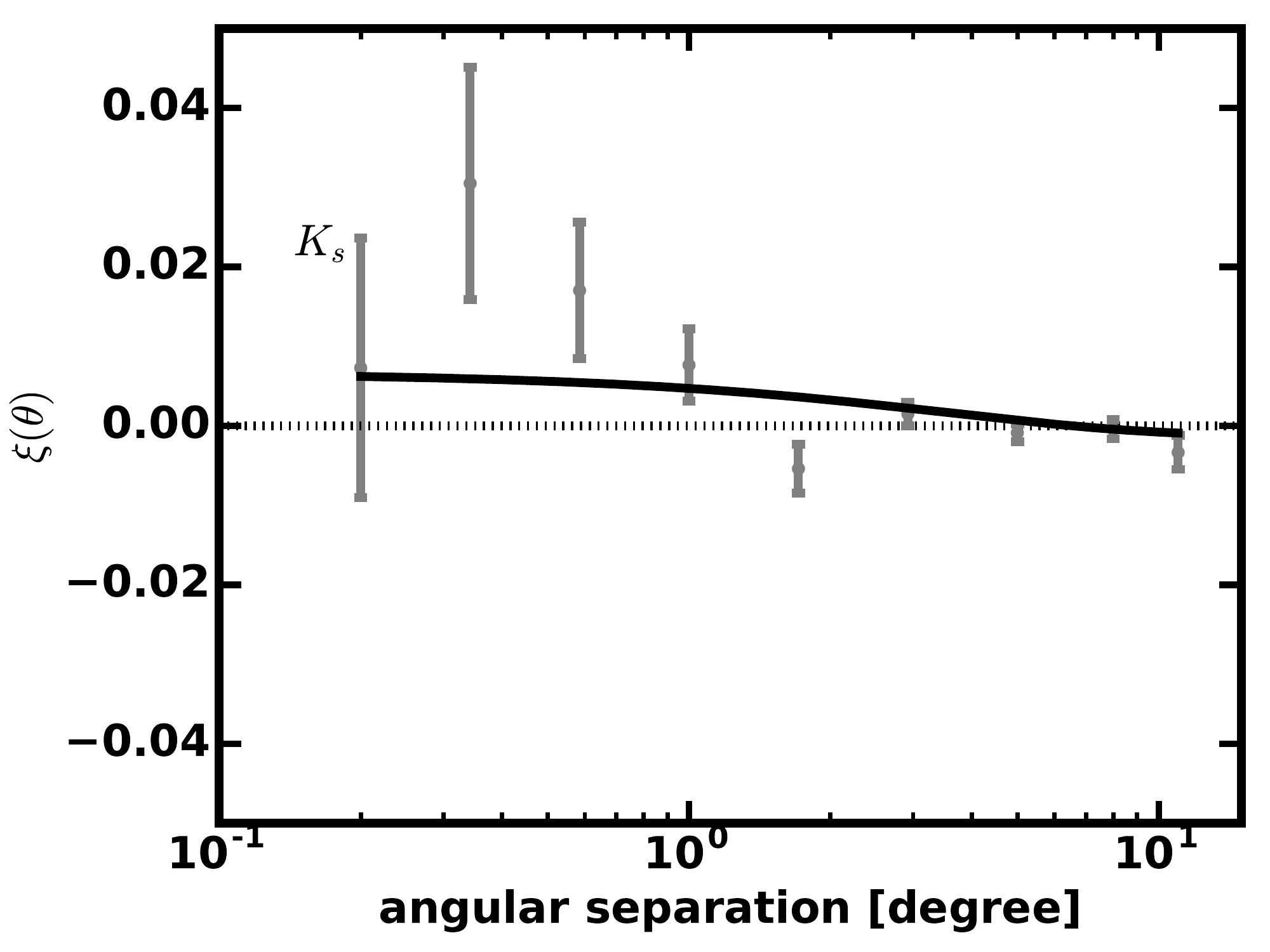}
}
\subfloat{
\centering
   \includegraphics[width=0.5\textwidth]{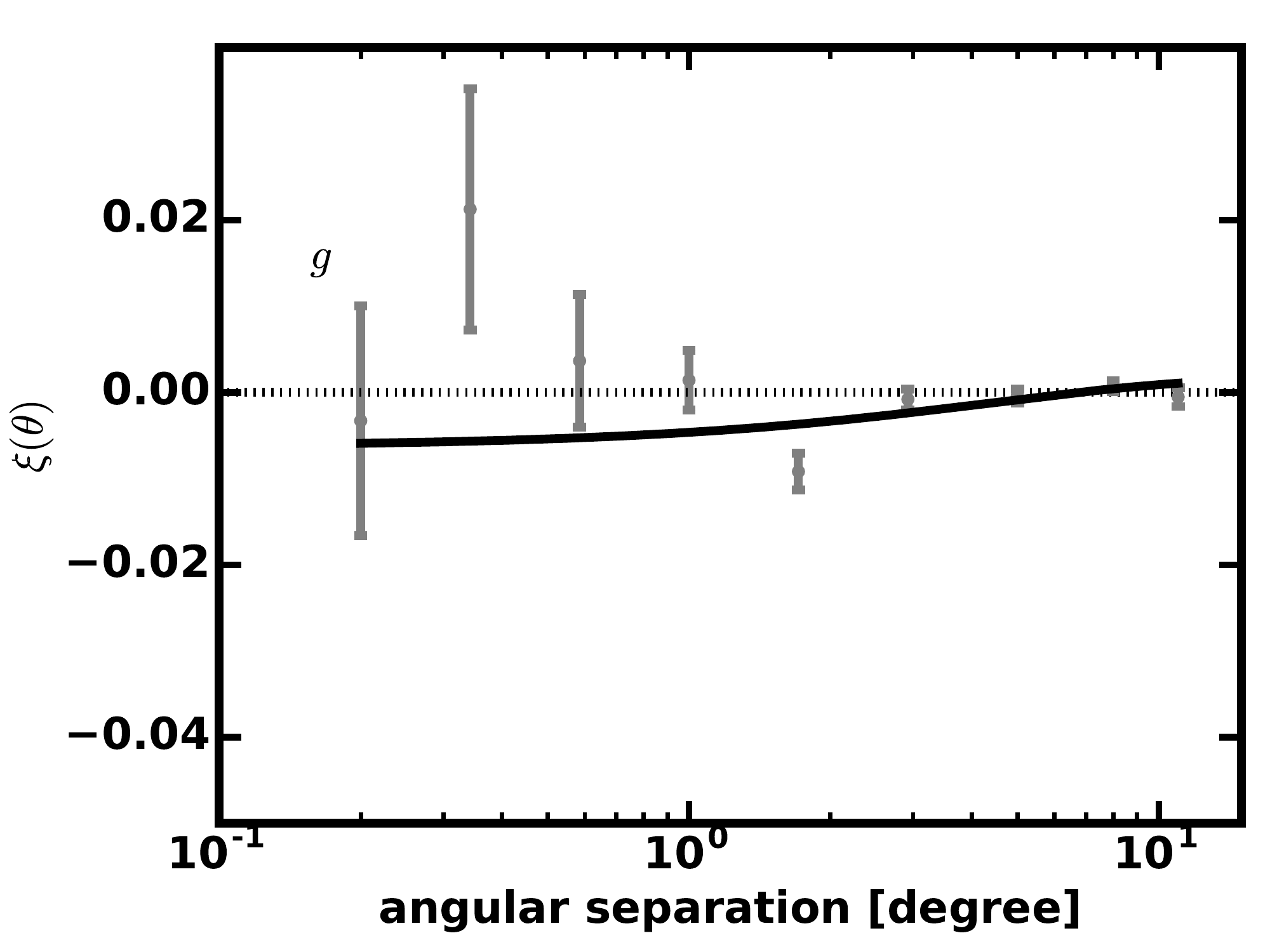}
}
\caption{Same as Figure~\ref{fig:ang_corr}, but for the observables on
  which asteroseismic parallax depends. Note the difference in scale
  for the $A_V$ panel.}
\label{fig:obs_corr1}
\end{figure*}

\begin{figure*}
\subfloat{
\centering
   \includegraphics[width=0.5\textwidth]{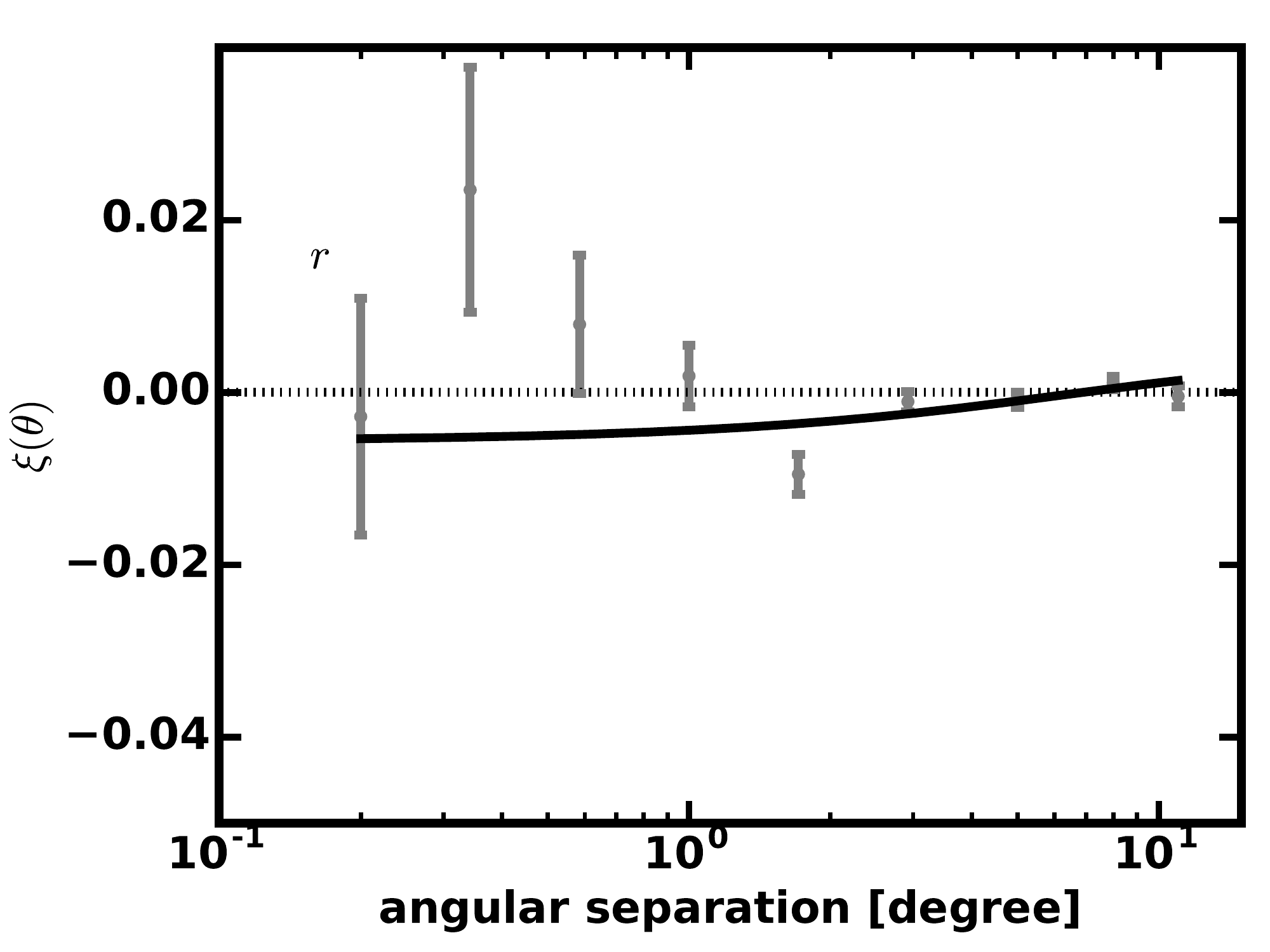}
}
\subfloat{
\centering
   \includegraphics[width=0.5\textwidth]{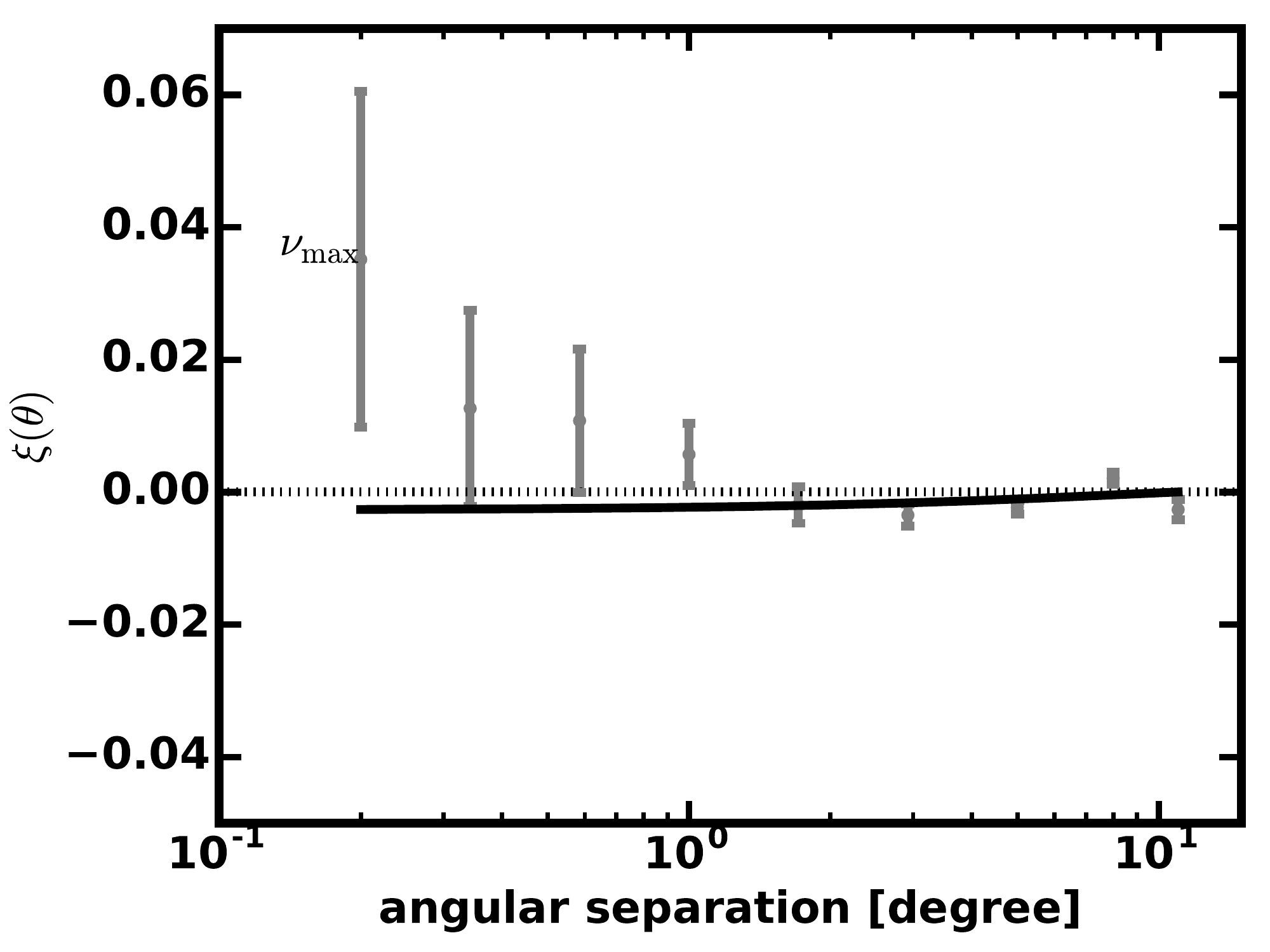}
}
\\
\subfloat{
\centering
   \includegraphics[width=0.5\textwidth]{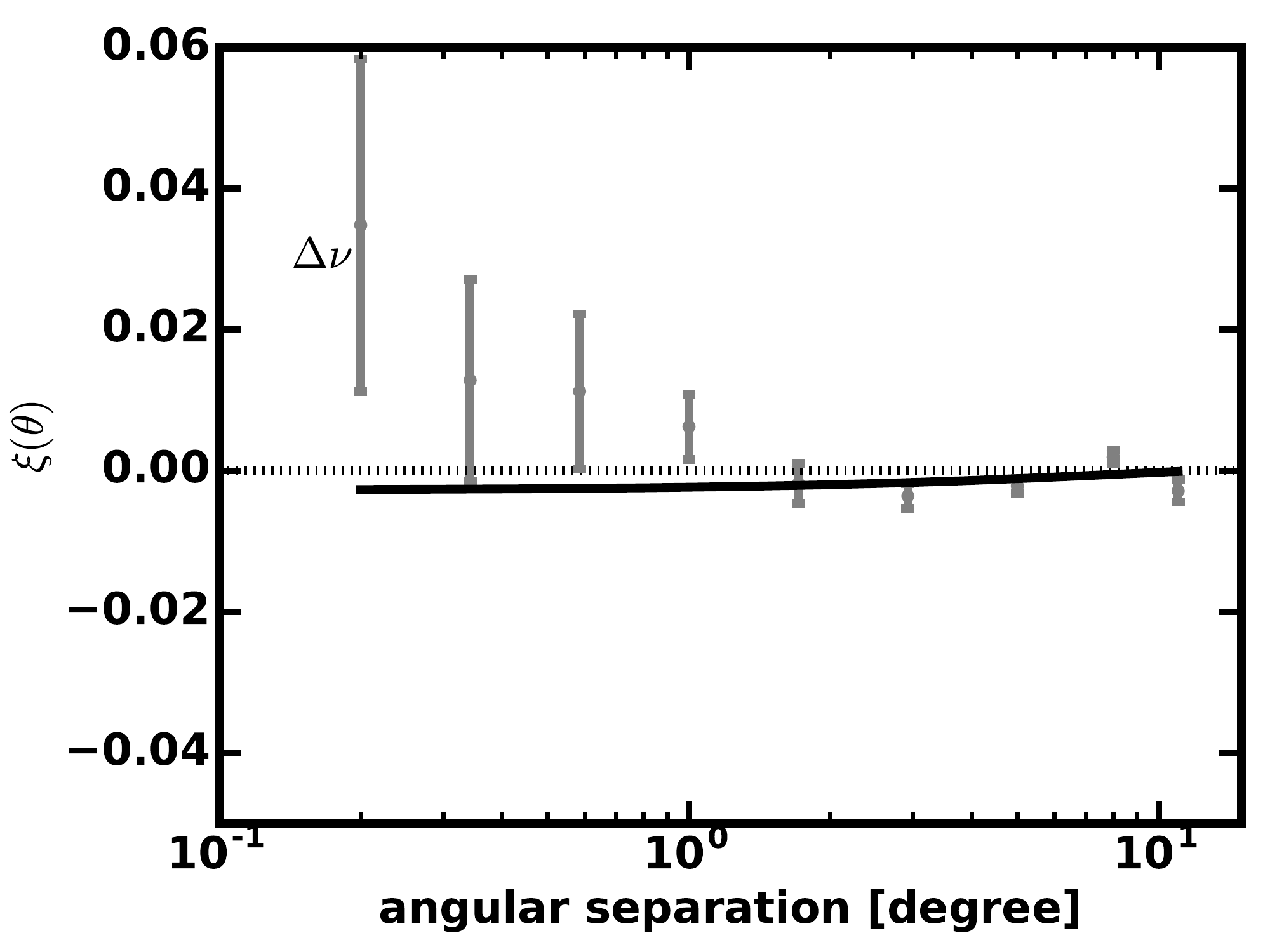}
}
\subfloat{
\centering
   \includegraphics[width=0.5\textwidth]{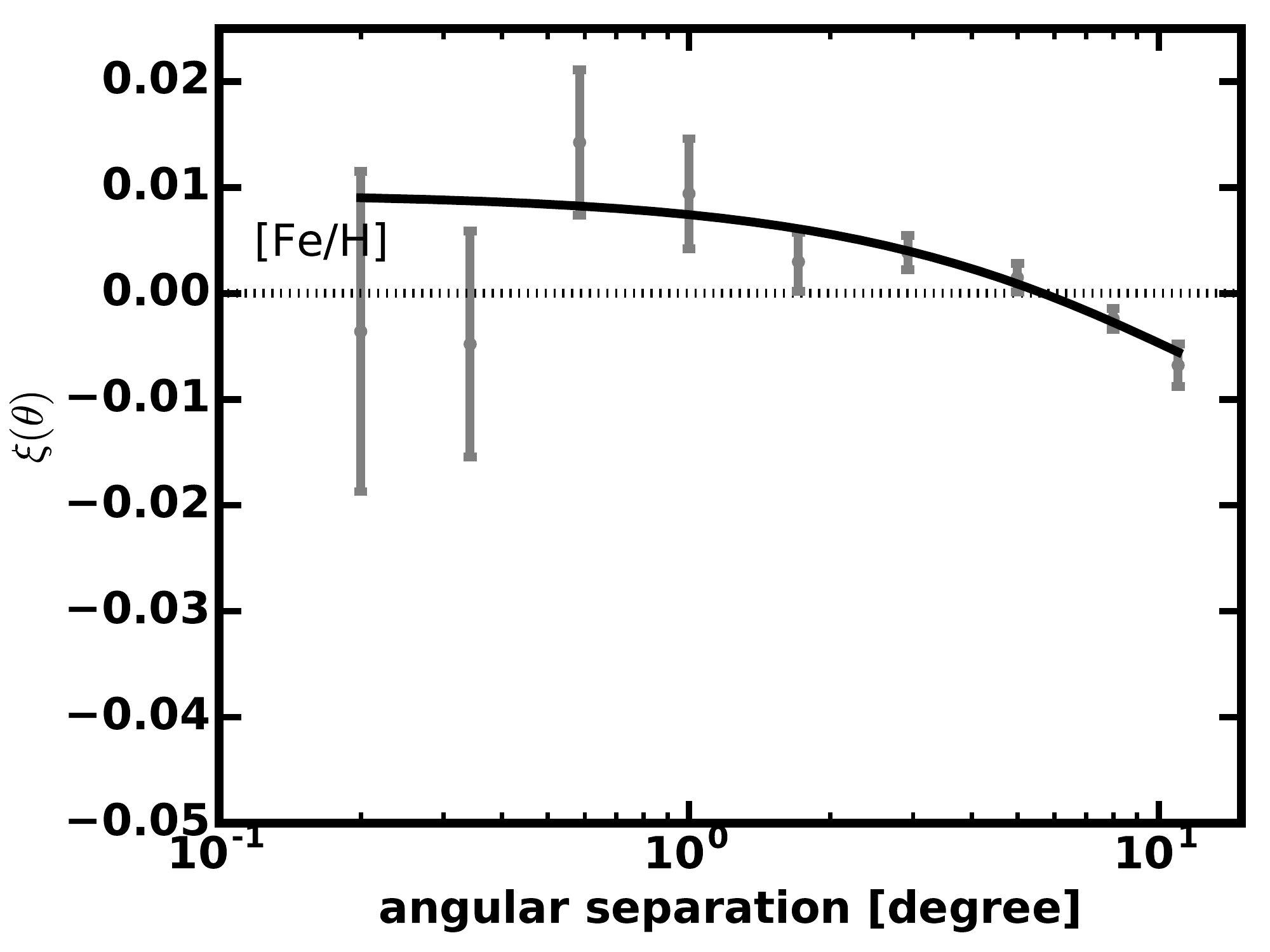}
}
\caption{Continued from Figure~\ref{fig:obs_corr1}.}
\label{fig:obs_corr2}
\end{figure*}

\begin{figure}[htb!]
\includegraphics[width=0.5\textwidth]{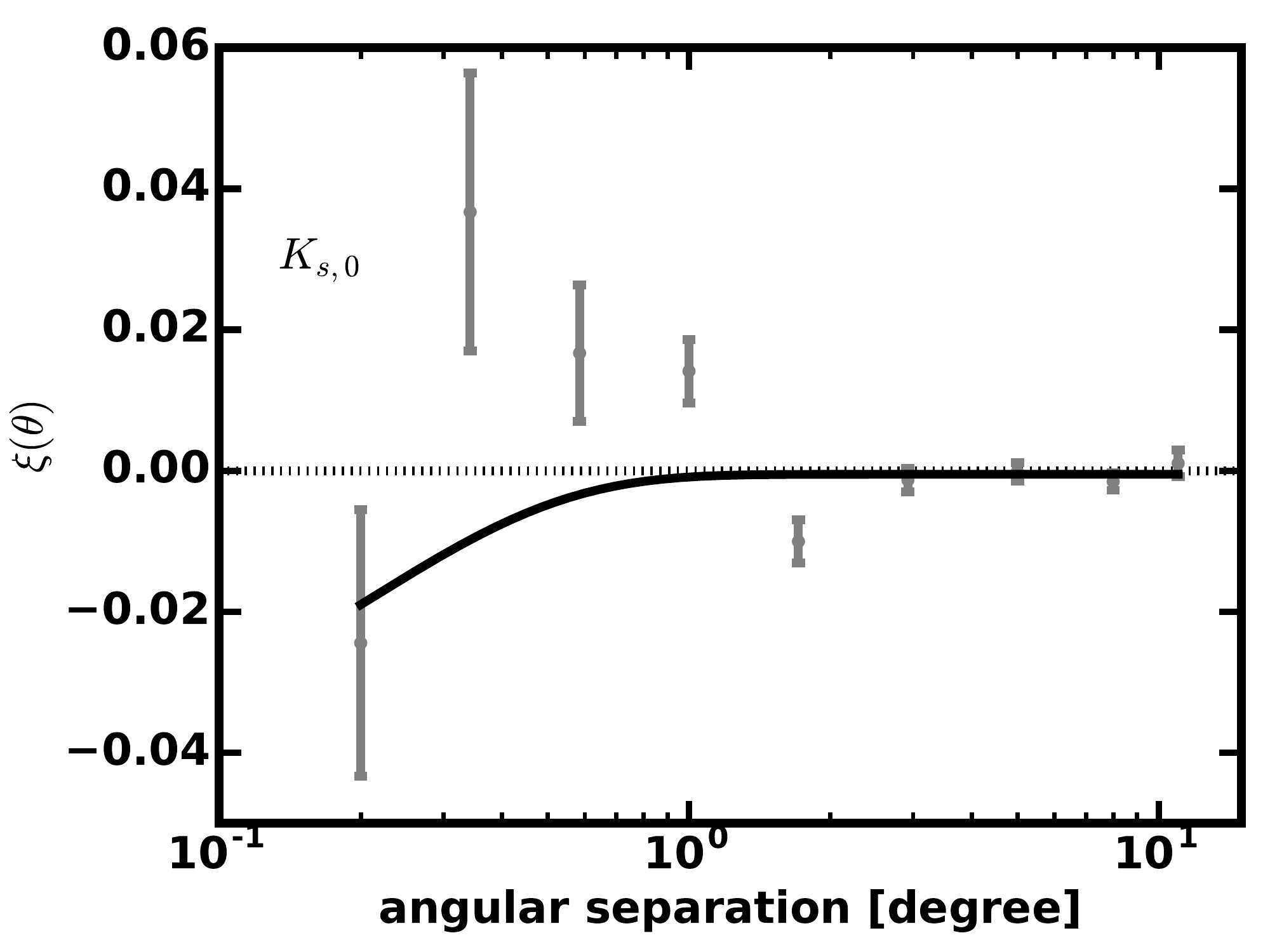}
\caption{Same as Figures~\ref{fig:obs_corr1} \&~\ref{fig:obs_corr2},
  but for de-extincted $K_{\mathrm{s}}$, $K_{\mathrm{s,\ 0}}$, calculated iteratively, as
  described in \S\ref{sec:plxastero}, and with asteroseismic
  parallax computed with a $K_{\mathrm{s,\ 0}}$ bolometric correction instead of
  $\fbol$ from the IRFM, and using
  $\teffapogee$ instead of $\teffirfm$. Refer to \S\ref{sec:obs} for details.}
\label{fig:kmag0}
\end{figure}

\subsection{Spatial correlation of observables}
\label{sec:obs}
It is possible that the observed spatial dependence of the quantity
$\plxtgas - \plxastero$ could be the result of spatial correlations in
the observables that enter in to the calculation of $\plxastero$. We
plot all the observable quantities that are used to compute
$\plxastero$, which are shown in
Figures~\ref{fig:obs_corr1} \&~\ref{fig:obs_corr2}. We have computed a
binned Pearson correlation coefficient for these quantities via a
modified version of Equation~\ref{eq:pearson}:
\begin{equation}
\xi (\theta) \equiv \frac{\sum_{i\neq j} (X_i - \langle X
  \rangle)_{\theta}(X_j - \langle X \rangle)_{\theta}}{\sqrt{((X_i -
    \langle X \rangle)^2)_{\theta}((X_j - \langle X \rangle)^2)_{\theta}}},
\label{eq:pearsonobs}
\end{equation}
where $\langle X \rangle$ is the average value of an observable
quantity, $X$, for the entire TGAS-APOKASC sample. The above quantifies, as a function of angular separation, how
correlated an observable quantity, $X$, is. Table~\ref{tab:obs} shows
that all observable
quantities except $A_V$ have negligible spatially-correlated
systematic errors, $\sigma_{\mathrm{sys}}$, based on fits to the
binned Pearson correlation coefficient, which are mapped to systematic error according to Equation~\ref{eq:cov}.

Unsurprisingly, there are non-negligible spatial correlations in our derived
extinctions. In addition to intrinsic spatial clustering of dust in
projection, the dust map in Figure~\ref{fig:dust} shows a region of
enhanced extinction in the {\it Kepler} field, which is spatially concentrated. Crucially, we do not
find significantly different results when we perform our
analysis only on the region in which the extinction is the highest
($\ell \lesssim 73\deg$).

Even a systematic spatial correlation at small angular scales
(less than $0.1$ degrees) at the level of $0.07$mag in
$A_V$ would at most translate as a $0.035$mas offset in parallax scales.
However, we infer a systematic offset between TGAS and asteroseismic
parallaxes of $0.127^{+0.010}_{-0.011}$mas near $0.1$deg for the
best-fitting model % STAT 
for the entire TGAS-APOKASC sample.\footnote{Here and for other parallax offsets quoted in mas
  in the paper, we assume the
  best-fitting model of Equation~\ref{eq:model2} fitted to the entire sample of
  red giants
  (`ALL'; this is the model plotted in Figure~\ref{fig:ang_corr} --- see also Table~\ref{tab:plx}) for the Pearson correlation coefficient,
$\xi(\theta)$. The correlation coefficient is translated into an absolute offset in mas
according to Equation~\ref{eq:cov}.} In other words,
the correlations in dust cannot account for the correlation we
see in TGAS parallaxes.

We performed additional tests to confirm that our 
result is not due to spatial correlations of extinction propagating into our asteroseismic parallaxes. Figure~\ref{fig:kmag0} shows the
binned Pearson correlation coefficient for de-extincted $K_{\mathrm{s}}$,
$K_{\mathrm{s,\ 0}}$. We note that there is statistically no
spatial correlation in $K_{\mathrm{s,\ 0}}$, as there is in $A_V$, which is due to
the negligible dust extinction in $K_{\mathrm{s}}$ ($A_{K_{\mathrm{s}}} \approx 0.1
A_V$). We therefore tested a single-band bolometric correction with
$K_{\mathrm{s}}$---instead of using $BVJHK_{\mathrm{s}}$, per our
fiducial IRFM method described in \S\ref{sec:methods}---and found that
it does not
remove the spatially-correlated parallax offset. We also confirm that
using the reddenings derived from stellar models
adopted in Huber et al. (in press)
instead of those from a dust map does not change our results.

\subsection{TGAS-asteroseismic parallax zeropoint}
\label{sec:zp_corr}
\cite{stassun&torres2016a} find that, when compared to 99 eclipsing
binaries from \cite{stassun&torres2016b} (which have parallaxes with
uncertainties of $\sim 200\mu$as), {\it Gaia} DR1 parallaxes are smaller by
$\sim 200\mu$as.
\cite{casertano+2017}, however, found no systematic offset above
$1\mu$as when comparing with 202 Galactic Cepheids with photometric
parallaxes. We find the absolute zeropoint correction suggested by \cite{stassun&torres2016a} to
over-correct the TGAS parallaxes by about $0.20 \pm 0.05$mas  (see % STAT
Figure~\ref{fig:zp}). Though this discrepancy in zeropoints between
TGAS and our parallax sample and that of \cite{stassun&torres2016a} is
likely because the zeropoint offset should be fractional and not
absolute, we take the
chance here to discuss potential zeropoint systematics in asteroseismic parallax.

One possible bias in asteroseismic parallaxes are systematics in the
scaling relations of Equations~\ref{eq:scaling1}--\ref{eq:scaling3}. The most evident
assumption in using asteroseismic scaling relations to determine stellar radii is the assumption
of homology in stellar structure relative to the Sun. Indeed, a growing body of
literature indicates that the $\dnu$ scaling relation can deviate by a
couple per cent when applied to derive RGB mean densities, compared to asteroseismic stellar models
\citep[e.g.,][]{white+2011,miglio+2012,guggenberger2016,sharma+2016}. Among
dwarfs, comparing
asteroseismic radii to radii from {\it Hipparcos} parallax and bolometric
flux \citep{silva_aguirre+2012} and to radii from interferometry
\citep{huber+2012} show agreement to within $5\%$. RGB radii
comparisons show mixed results
\citep[see][]{huber+2012,frandsen+2013,gaulme+2014,brogaard+2016,gaulme+2016}.

We investigate the validity of scaling relations in Huber et
al. (in press) by comparing to {\it Gaia} parallaxes by averaging
over the entire {\it Kepler} field of view. The effect of
spatially-correlated offsets in asteroseismic and TGAS parallaxes at
the level indicated in this work does not affect the result that
asteroseismic radii are consistent with {\it Gaia} radii at the $5$ per
cent level. We also find
evidence for $\dnu$ corrections proposed by \cite{sharma+2016} improving global agreement between the
two radii scales.

In light of evidence for red giant $\dnu$ scaling relation
corrections, we conservatively apply the $\teff$- and $\feh$-dependent
corrections that \cite{sharma+2016} propose to our APOKASC $\dnu$ values.
Such corrections do not significantly affect $\logg$ calculations
\citep{hekker+2013}, and the significance of our result is not affected by the choice of whether or not to
apply $\dnu$ corrections.

Even if the asteroseismic scale is biased with respect to the TGAS parallax
scale (which would manifest as a zeropoint offset), the spatial
correlation that we retrieve is still valid. Any required corrections to the
asteroseismic scaling relations (Equations
\ref{eq:scaling1}--\ref{eq:scaling3}) will only be spatially-dependent
insofar as the observable quantities that enter in to them (i.e.,
$g$, $r${, $J$, $H$, $K_{\mathrm{s}}$}, $\numax$, $\dnu$, $\feh$, $\teff$) are significantly
spatially correlated. We demonstrate in 
\S\ref{sec:obs} that no significant spatial correlations exist in
these quantities, except in $A_V$. Importantly, we have confirmed that the result is insensitive to
extinction corrections, which are, in fact, spatially-correlated on
the angular scales investigated in this paper.

\subsection{Gaia systematics}
\label{sec:gt10}
If the observed difference in TGAS and asteroseismic parallax scales is not
due to spatial correlations of asteroseismic parallaxes themselves, then we
interpret them as spatially-correlated errors in TGAS
parallaxes. There are a few reasons to come to this conclusion. As
discussed in \cite{lindegren+2016}, for instance, the attitude model of the
astrometric solution does not have a high enough temporal resolution
to remove small time scale attitude changes. As
a result, the {\it Gaia} team expects that spatial correlations on scales
of a few degrees and less are a result of unmodeled, correlated
attitude changes on time scales of minutes,
which translate into spatial correlations of degrees and less.

\begin{figure}[htb!]
\centering
\includegraphics[width=0.5\textwidth]{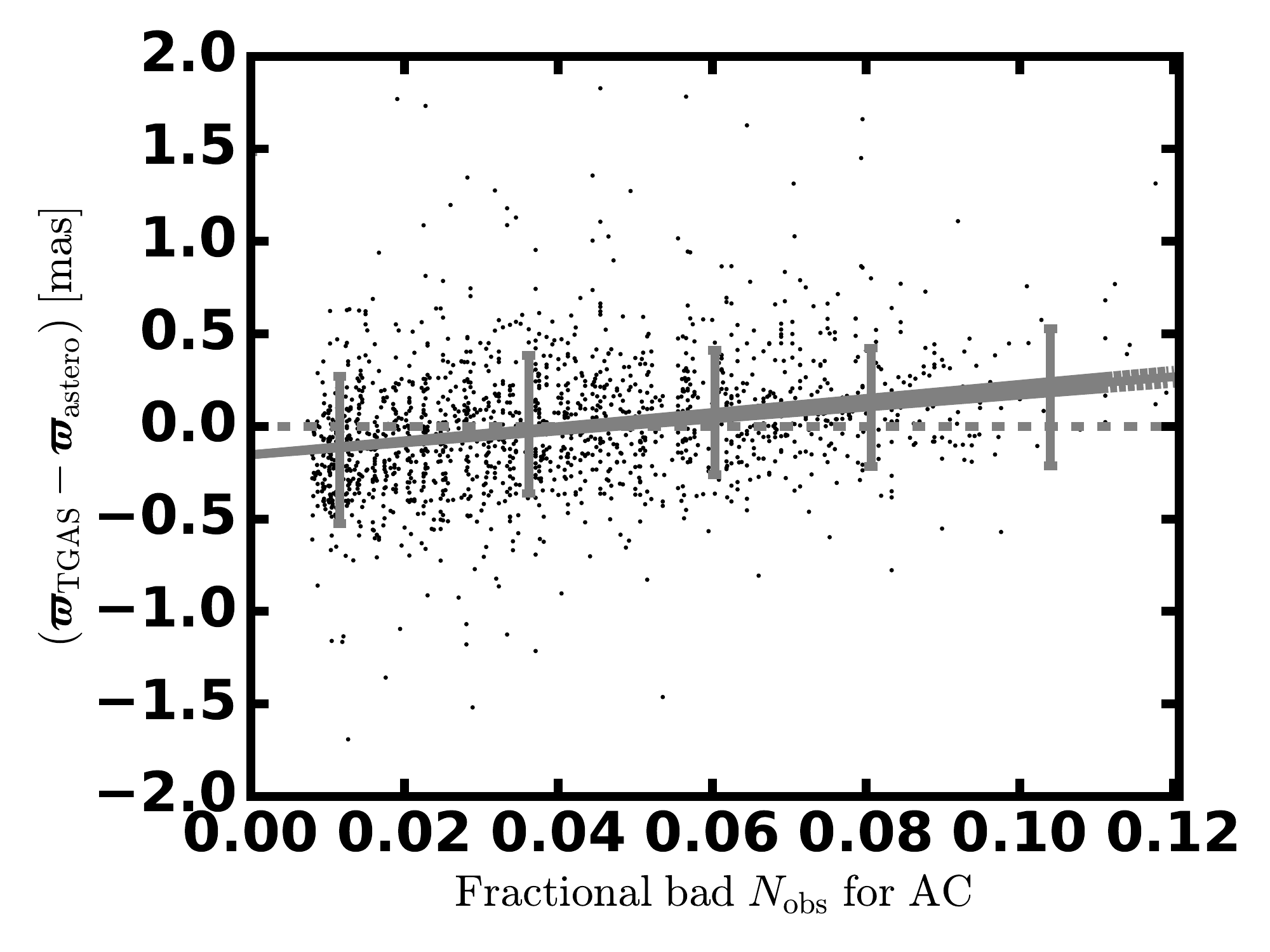}
\caption{Difference in TGAS and asteroseismic parallax for each star
  in the TGAS-APOKASC sample, as a
  function of the ratio of the number of across-scan measurements flagged as `bad'
  to the total number of across-scan measurements for each star. The
  straight line is the best-fitting trend, with 1-sigma slopes shown as
  dashed lines. The standard deviation of parallax difference in bins
  of the number of `bad' across-scan measurements are shown as 
  error bars.}
\label{fig:ac}
\end{figure}

Though we do not have access to the astrometric solution model to
independently demonstrate that observed TGAS-asteroseismic parallax offsets are a result of systematic errors
in the astrometric solution, we can make
inferences based on the published DR1 data. In
particular, we show in Figure \ref{fig:ac} that the difference in
TGAS and asteroseismic parallax scales correlates significantly with the
fraction of `bad'
across-scan direction observations to total across-scan direction
observations. Taking this metric as a proxy for the
uncertainty in the across-scan measurement, the correlation
corroborates a note in
\cite{lindegren+2016} indicating that the across-scan direction
measurement error changes parallax solutions in a systematic way, for
unknown reasons.

\section{Conclusion}
\label{sec:conclusion}
We have independently validated and have quantified spatially-correlated errors in
TGAS parallaxes, as predicted by the {\it Gaia} team. Our result complements warnings in the {\it Gaia} DR1
documentation that there exist systematic uncertainties of amplitude
comparable to the statistical uncertainties. For convenience and
comparison to future work, we have provided a characteristic scale and
amplitude for the spatial correlations: an error of $0.059^{+0.004}_{-0.004}$mas on scales
of $0.3$deg, which decreases for larger scales to become
$0.011^{+0.006}_{-0.004}$mas at % STAT
$8$deg. A covariance matrix for the correlated errors in parallax
may be computed via Equation~\ref{eq:cov}, using either of the
models fit to the observed spatial correlation signal, $\xi(\theta)$, which are
provided in Table~\ref{tab:plx}. For any pair of stars, $i$ and
$j$, separated by angular distance, $\theta$, their respective entry in a covariance matrix, $\sigma^2_{ij}$, would
read $\sigma_{ij}^2 = \xi(\theta)\sigma^2$, where $\sigma$ is the
statistical error on TGAS parallaxes.

We have done several checks on our result, which is robust to:

\begin{enumerate}
\item the dust prescription that is used --- without
significant differences in the observed spatial correlation in
parallax error when omitting the region of the Kepler field of view
most affected by dust or when using a stellar model extinction
instead of a dust map extinction;
\item the evolutionary status of the stars used to calculate
  asteroseismic parallax, with both first ascent red giant branch
   and red clump parallaxes indicating the same spatially-correlated
  parallax offset with respect to TGAS parallaxes;
\item whether or not a $BVJHK_{\mathrm{s}}$ bolometric correction is used to
  compute asteroseismic parallax or a $K_{\mathrm{s}}$-band bolometric correction
  is used;
{
\item the temperature scale used -- whether it be the spectroscopic
  $\teffapogee$ scale or the IRFM scale, $\teffirfm$;
\item and whether or not $\dnu$ corrections are applied to the asteroseismic scaling relations.
}
\end{enumerate}

At this point, we % STAT
cannot test the possibility of correlations on scales larger than
$10$deg due to the $\sim10\deg \times 10\deg$ spatial extent of the
{\it Kepler} field of view. Future work could quantify spatial correlations on larger scales using, e.g., K2 asteroseismology, which would yield parallaxes for objects separated by the largest angular scales. We encourage the use of the spatial
covariance functional form when computing quantities that depend on
TGAS parallaxes, especially in light of the delay of {\it Gaia} DR2 to
April 2018.

\acknowledgments
We would like to thank A. Gould, C. Kochanek, B. Wibking, and A. Salcedo for
useful discussions. DH acknowledges support
by the Australian Research Council's Discovery Projects
funding scheme (project number DE140101364) and support by the
National Aeronautics and Space Administration under Grant NNX14AB92G
issued through the {\it Kepler} Participating Scientist
Program. Funding for the {\it Kepler} Mission
is provided by NASA's Science Mission Directorate.

This research made use of the cross-match service provided by CDS,
Strasbourg. This publication makes use of data products from the Two Micron
All Sky Survey, which is a joint project of the University
of Massachusetts and the Infrared Processing and Analysis Center/California Institute of Technology, funded by
the National Aeronautics and Space Administration and
the National Science Foundation.

Funding for the Sloan Digital Sky Survey IV has been
provided by the Alfred P. Sloan Foundation, the U.S.
Department of Energy Office of Science, and the Participating Institutions. SDSS acknowledges support and
resources from the Center for High-Performance Computing at the University of Utah. The SDSS web site
is www.sdss.org. SDSS is managed by the Astrophysical Research Consortium for the Participating Institutions of the SDSS Collaboration including the Brazilian
Participation Group, the Carnegie Institution for Science, Carnegie
Mellon University, the Chilean Participation Group, the French
Participation Group, Harvard Smithsonian Center for Astrophysics, Instituto de Astrofs{\'i}ca de Canarias, The Johns Hopkins University,
Kavli Institute for the Physics and Mathematics of
the Universe (IPMU) / University of Tokyo, Lawrence
Berkeley National Laboratory, Leibniz Institut f{\"u}r Astrophysik Potsdam (AIP), Max-Planck-Institut f{\"u}r Astronomie (MPIA Heidelberg), Max-Planck-Institut f{\"u}r Astrophysik (MPA Garching), Max-Planck-Institut f{\"u}r Extraterrestrische Physik (MPE), National Astronomical
Observatories of China, New Mexico State University,
New York University, University of Notre Dame, Observat{\'o}rio Nacional / MCTI, The Ohio State University,
Pennsylvania State University, Shanghai Astronomical
Observatory, United Kingdom Participation Group, Universidad Nacional
Aut{\'o}noma de M{\'e}xico, University of Arizona, University of Colorado
Boulder, University of Oxford, University of Portsmouth, University of
Utah, University of Virginia, University of Washington, University of Wisconsin, Vanderbilt University, and Yale University.

\appendix
\section{Alternate formulations of spatial correlation}
Here, we present representations of spatially-correlated parallax offsets that are complementary to the Pearson correlation
coefficient presented in the text. We prefer the Pearson correlation
coefficient formulation because it is directly mappable to a
covariance function, but we include the following representations for completeness.
\subsection{Angular correlation function}

\begin{figure}[htb!]
\centering
\includegraphics[width=0.5\textwidth]{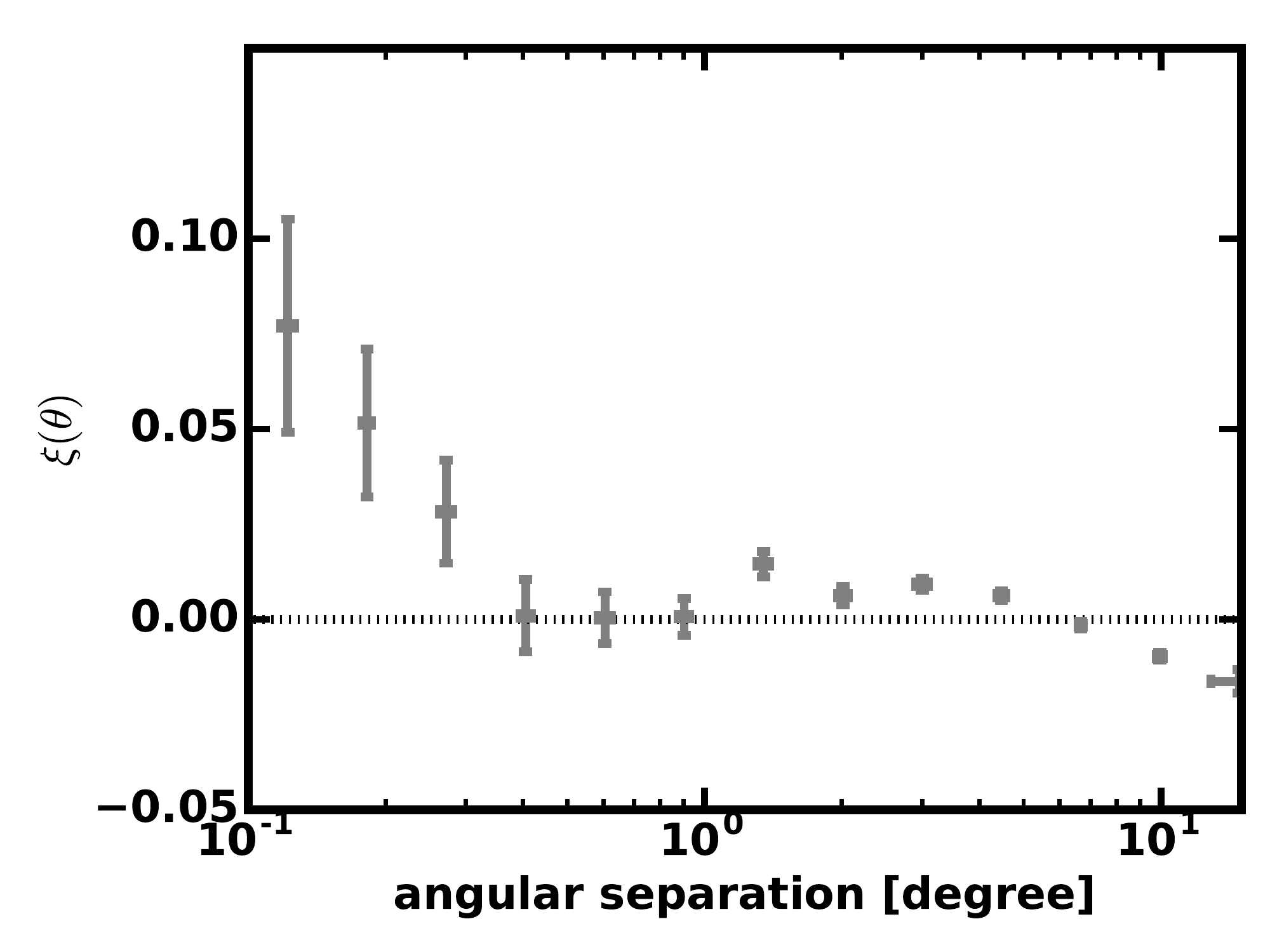}
\caption{The angular correlation function of the quantity $(\plxtgas
  - \plxastero)/\sqrt{(\sigma_{\plxtgas}^2 +
    \sigma_{\plxastero}^2)}$, which will be positive for offsets that are positively correlated (i.e., differences between TGAS and asteroseismic parallaxes are in the same sense for pairs of stars at a given angular scale), zero in the absence of spatially-correlated parallax offsets, and
  negative for anti-correlated TGAS parallax offsets (i.e., differences between TGAS and asteroseismic parallaxes are in opposite senses for pairs of stars at a given angular scale). Grey points are
  the observed angular correlation function values, with error bars assuming
  Poisson statistics; the black dashed line indicates
  a null correlation. See the Appendix for details.}
\label{fig:ang_corr_treecorr}
\end{figure}

The angular correlation function, often used in cosmological contexts,
may be calculated using the Landy-Szalay
estimator \citep{landy&szalay1993}:
\[
\xi(\theta) = \frac{\langle\mathrm{DD}\rangle_{\theta} - 2\langle\mathrm{DR}\rangle_{\theta}+ \langle\mathrm{RR}\rangle_{\theta}}{\langle\mathrm{RR}\rangle_{\theta}},
\label{eq:ls}
\]
where $D$ in our case refers to an observed value of the quantity
$(\plxtgas - \plxastero)/\sqrt{(\sigma_{\plxtgas}^2 +
  \sigma_{\plxastero}^2)}$ for a star, and $R$ refers to a sample drawn
from the observed values of the normalized parallax difference, but
with positions drawn randomly from within the {\it Kepler} field of view,
making use of
\texttt{K2fov}\footnote{\href{https://github.com/mrtommyb/K2fov}{https://github.com/mrtommyb/K2fov}}
\citep{mullally_barclay&barentsen+2016}; $\langle \rangle_{\theta}$ represents the expected
value of that quantity for pairs of points separated by angular
distance $\theta$. We compute this statistic with \texttt{TreeCorr}\footnote{\href{https://github.com/rmjarvis/TreeCorr}{https://github.com/rmjarvis/TreeCorr}}
\citep{jarvis_bernstein&jain2004}. Error bars for each angular bin are assigned based on Poisson
statistics. This statistic is widely-used in cosmology to compute
correlation functions. Although complementary to, the angular
correlation coefficient will in general not be equivalent to the
Pearson correlation coefficient. However, it does explicitly account for stochasticity in the spatial
distribution of the TGAS-APOKASC sample.  Results using this approach
are qualitatively similar, as seen in comparing
Figures~\ref{fig:ang_corr} \&~\ref{fig:ang_corr_treecorr}.

\subsection{Binned absolute difference}
\begin{figure}[htb!]
\centering
\includegraphics[width=0.5\textwidth]{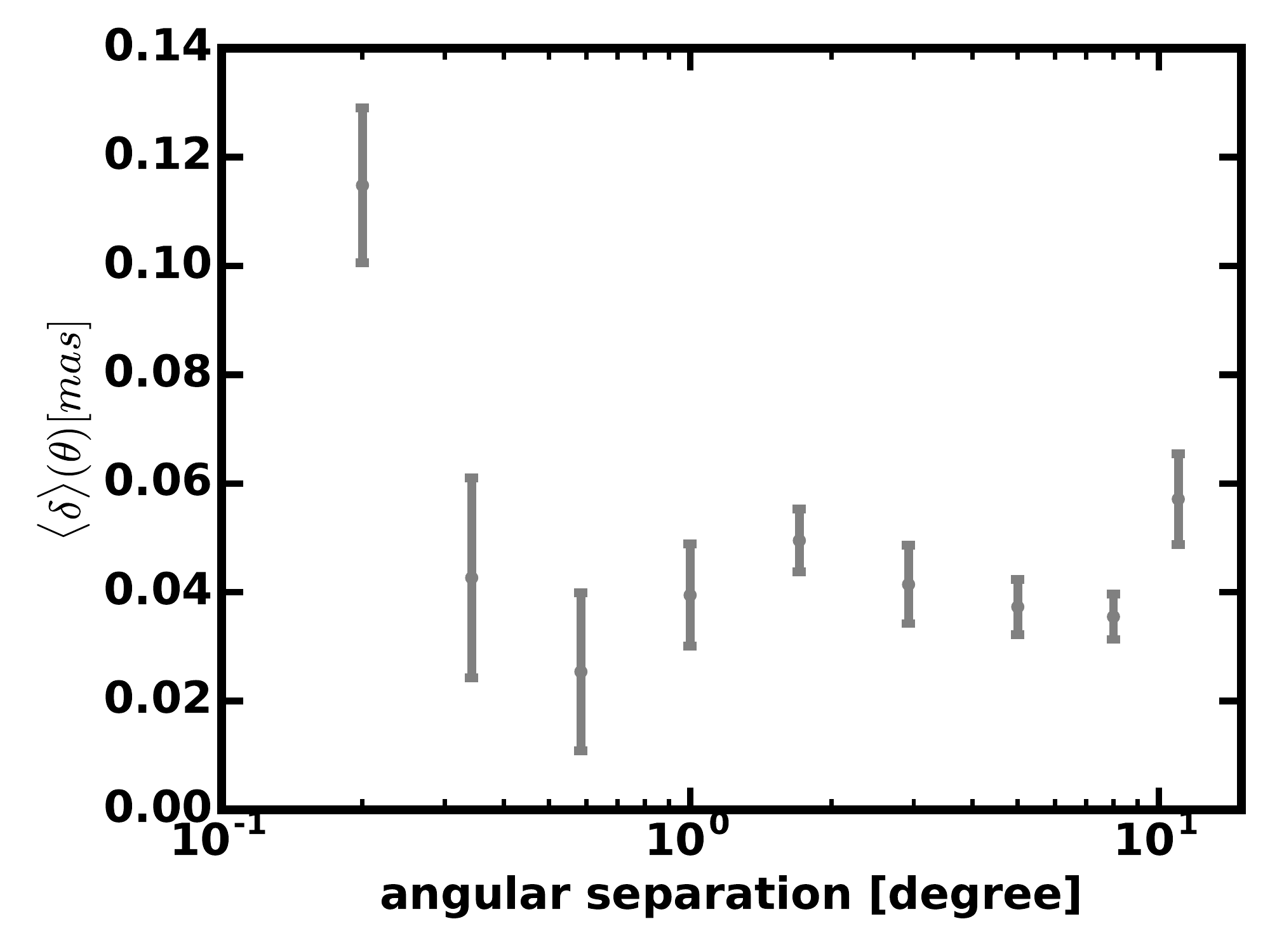}
\caption{A binned absolute difference in parallax scales shows excess
  difference in the scales for angular scales less than a few
  degrees, offering an alternate representation of the main result of
  spatially-correlated offsets in asteroseismic and TGAS parallax
  scales (see Figure~\protect\ref{fig:ang_corr}). See the Appendix for
details.}
\label{fig:abs}
\end{figure}

Most intuitive is the simple measure
\[                                                                                                            
\langle \delta \rangle (\theta) = \sqrt{\langle|(\plxtgasi - \plxasteroi)(\plxtgasj - \plxasteroj)|\rangle}_{\theta},
\label{eq:ls}                                                                                                 
\]
which is a measure of the absolute difference in the parallax scales
computed by binning pairs of stars, $i$ and $j$, separated by an
angular distance, $\theta$. This scale will not necessarily be the
same as the scale we present in the text, and in particular
is insensitive to the sign of the (anti-)correlation. Figure~\ref{fig:abs} shows this
measure { for the TGAS-APOKASC sample}. 
\clearpage                         
\newpage                          

\bibliography{zinn_submit}
%\begin{thebibliography}
%\end{thebibliography}
\label{lastpage}
\end{document}